\documentclass[prd,twocolumn,showpacs,preprintnumbers,aps,nofootinbib]{revtex4-1}

\usepackage{graphicx}
\usepackage{dcolumn}
\usepackage{bm}
\usepackage{amsmath,amssymb,amsfonts}
\usepackage{latexsym}

\usepackage{color}

\def\nn    {\nonumber}

\newcommand{\be}{\begin{equation}}
\newcommand{\ee}{\end{equation}}
\newcommand{\bea}{\begin{eqnarray}}
\newcommand{\eea}{\end{eqnarray}}

\begin{document}


\title{
Decadal Mission for the New Physics Higgs/Flavor Era
}

\author{Wei-Shu Hou}

\affiliation{
Department of Physics, National Taiwan University, Taipei 10617, Taiwan}
\bigskip


\begin{abstract} 
The LHC has not discovered any New Physics beyond 
the anticipated $h(125)$ boson, and new ideas abound for 
out-of-the-box searches, or Effective Field Theories with multi-TeV cutoff. 
But, have we exhausted dimension-4 operators involving sub-TeV 
particles that are {\it not} exotic (non-XLP)? 
We advocate the existence of an extra Higgs doublet that 
carry extra Weinberg couplings, where {\it emergent} 
mass-mixing hierarchies {\it and} alignment 
have hidden their effects very well so far.
${\cal O}(1)$ Higgs quartics can induce first order electroweak phase transition
and imply sub-TeV $H, A, H^\pm$ spectrum. 
The extra Weinberg couplings, led by $\rho_{tt}$ and $\rho_{tc}$ 
at ${\cal O}(1)$, can drive electroweak baryogenesis, 
while $|\rho_{ee}/\rho_{tt}| \propto \lambda_e/\lambda_t$, 
the ratio of usual electron and top Weinberg couplings, can tame electron EDM.
Finding these extra Higgs bosons via $cg \to tH/A \to tt\bar c$, $tt\bar t$ 
and $cg \to bH^+ \to bt\bar b$ processes
 (and those allowed by Higgs boson splittings) 
at the LHC, and pushing the flavor frontier to break the flavor code, 
would usher in a new Higgs/flavor era.
A new scale at 10--20 TeV, possibly related to 
the Landau pole of the scalar sector, may emerge.
\end{abstract}

\maketitle


Physics is an empirical science. 
For the Higgs and ``Flavor'' era unfolding before us, we have to thank
the large LHC experiments and Belle~II, plus smaller, more specialized 
experiments at the kaon, muon and electric dipole moment (EDM) frontiers.

\section{\boldmath
Prelude: $e$, $m_e$, $\mu$, the Universe 
}

Modern physics share a common heritage. 
Electrodynamics, completed by Maxwell, is still the template of dynamics.
Thomson's electron, a hundred and twenty four years and going, remains 
fundamental: we still have not resolved any structure it may have.
Rutherford discerned that Becquerel's radioactivity 
separates into $\alpha$, $\beta$ and $\gamma$ rays.
Demonstrating $\alpha = {\rm He}^{++}$, and lucking out that 
it is the tightest bound nucleus, he put it to good use and 
made the monumental discovery of the {\it nucleus} itself;
and $\gamma$ rays unraveled as but an energetic repeat of atomic spectra.
However, the $\beta$ ray, energetic electron emission from the nucleus, 
turned out to be the manifestation of a new fundamental interaction
heretofore unbeknownst to man: the weak interaction.

With the backdrop of QED and the key provided by Chadwick's neutron, 
Fermi took up Pauli's audacious hypothesis of the nearly massless neutrino to 
formulate his theory of the weak interaction as 
a product of  fermion bilinears, e.g. $(\bar pn)(\bar \nu e)$. 
In turn, this inspired Yukawa's short range nuclear force 
by massive pion exchange, which started nuclear physics.
But it was the development of ultra-low temperature techniques 
by Kamerlingh Onnes, resulting in his discovery of superconductivity, 
that provided the seed of inspiration for the later development of 
the Standard Model (SM) of particle physics, 
namely spontaneous symmetry breaking (SSB) 
as {\it the origin of mass}. Who would have thought!

With atomic spectra explained by quantum mechanics, 
one could already ask: 

\quad ``Why is the coupling constant $e$ real?''

\quad ``What gives rise to $m_e$?'' \;\,\ \ [and me]

\noindent With Dirac's relativistic theory of the electron
settled by Anderson's discovery of $e^+$, the anti-electron, 
who went on to discover the muon with $m_\mu \gg m_e$ 
but otherwise electron-like, two further questions arose:

\quad ``Where have all the $e^+$ gone?'' 

\quad ``Who ordered the muon?"

\noindent where the latter was quipped by Rabi after 
Powell's demonstration of the $\pi$--$\mu$--$e$ decay chain.

If $\alpha$ bombardment utilized radioactivity provided by Nature,
a Rutherford scattering experiment in the late 1960s utilized man-made 
high energy electron {\it beams} to probe inside the proton and discover 
another layer of constituents, the ``reality'' of $u$ and $d$ quarks, 
which stand with the electron as fundamental fermions.
The Rabi question soon proliferated into the ``flavor'' problem:
the electron, neutrino\footnote{
The neutrinos, though having mass, are so light 
such that they belong in their own category, which we will not cover.
} 
and $u$ and $d$ quarks each come in three copies in Nature, 
increasing in mass.
Of these, the $s$ quark echoes the muon, while the
$\tau$ lepton and the $c$, $b$ and $t$ quarks were 
discovered since November 1974.
The latter four are called ``heavy flavors'' because they are supra-GeV in mass.
But $\mu$ and $s$ share property features with these heavy flavors,
even though $m_\mu, \, m_s \sim 0.1$~GeV.
The span in mass, from $m_e \simeq 0.511$~MeV to $m_t \simeq 173$~GeV,
is simply {\it staggering}.

With the development of gauge field theory, we now understand why $e$ is real,
but morphs the question into: 

\quad ``Are there complex couplings?''

\noindent Together with how $m_f$ arises ($f$ now covering 9 charged fermions),
where has antimatter gone, and the flavor problem, these are the issues we wish to
touch upon in this targeted Review.

{\it \textbf{Start of Particle Physics}.---} \
Anderson's discovery of ``the positive electron''~\cite{Anderson:1933mb}, 
or anti-electron $e^+$, settled all worries of Dirac, 
but brought in the issue as big as the Universe itself:

\quad ``Where have all the antimatter gone?'' 

\noindent Since luminal matter is dominated by baryons, the fact that 
we see baryons but no equivalent amount of anibaryons
is called the Baryon Asymmetry of the Universe (BAU) problem.
Remarkably, Anderson also discovered 
the muon~\cite{Neddermeyer:1937md} a few years later. 
But because of issues with the nature of cosmic rays 
and confusion with Yukawa's pion, things were not
clarified until the $\pi$--$\mu$--$e$ decay chain 
was established~\cite{Lattes:1947mx} by Powell, 
which coined $\pi$ and $\mu$.
Pontecorvo then stepped in to show 
the absence~\cite{Hincks:1948vr} of $\mu \to e\gamma$, 
hence $\mu \neq e^*$, i.e. it is not an excited electron.

The discovery of the positron and the muon, which are fundamental fermions
to this day, can be viewed as the start of modern particle physics.

{\it \textbf{\boldmath From $CP$ Violation to Baryogenesis}.---} \
After Lee and Yang's parity ($P$) violation suggestion, which was quickly
confirmed by Madame Wu, for 6--7 years, particle physicists embraced 
the conservation of $CP$, the product of 
charge conjugation $C$ and $P$, in weak interactions. 
However, Cronin and Fitch showed~\cite{Christenson:1964fg} experimentally 
that $CP$ is violated in $K_L^0 \to 2\pi$ decays, albeit minutely.
Armed with this and the concurrent observation~\cite{Penzias:1965wn} 
of the Cosmic Microwave Background (CMB),
i.e. evidence of a hot early Universe, the Big Bang, 
during a lull of the nuclear arms race, the father of the Soviet H bomb, 
Sakharov, proposed~\cite{Sakharov:1967dj} his conditions for BAU:

 1) Baryon Number Violation (BNV); 

 2) ($C$ \&) $CP$ Violation; 

 3) Deviation from Equilibrium.

\noindent The BNV insight, ahead of its time, was injected by Sakharov.

{\it \textbf{\boldmath Weak Interaction and $W/Z$ Bosons}.---} \
The muon decays weakly, $\mu \to e\nu_\mu\bar\nu_e$, which follows Fermi theory,
and it was shown experimentally that $\nu_\mu \neq \nu_e$.
With parity violation pointing the way to the $V-A$ theory,
it became clear that weak decay occurs via {\it massive} vector boson exchange.
The SU(2)$_L \times$U(1) gauge group was proposed~\cite{Glashow:1961tr},
where $L$ corresponds to left-handedness of weak processes.
{\it Somehow} the $W$ and $Z$ bosons gained mass, and only
the U$_Q$(1) symmetry of QED is left, with $m_\gamma = 0$ as observed.
It was the exchange of virtual photons in Deeply Inelastic Scattering
that unraveled the point-like partons, the $u$ and $d$ quarks, inside the nucleon.
The observed ``asymptotic freedom'' led to the SU(3) group of
Quantum Chromodynamics (QCD), the exquisite 
non-Abelian gauge dynamics that underlies nuclear binding.
The gauge symmetry remains unbroken, so $m_g = 0$,
although confinement means the gluon has an effective, dynamical mass.

The renormalizability, or control of divergences of non-Abelian gauge theories,
was a big issue, but was resolved by 't Hooft and Veltman, 
without or with SSB.

{\it \textbf{Mei\ss ner Effect and SSB}.---} \
With the mention of the Mei\ss ner Effect, discovered in 1933~\cite{Meissner:1933},
levitated high speed trains come to mind: 
magnetic fields cannot penetrate a superconductor.
Though it was explained phenomenologically by the London theory,
it took the insight~\cite{Anderson:1962} of (Phil) Anderson to see through the cloud:
the massless Goldstone boson of spontaneous breaking of charge symmetry
(by condensation of Cooper pairs)
became the longitudinal component of the photon, 
hence the photon gains mass inside the superconductor! 
And thus explains the attenuation length. This is exactly 
the Brout-Englert-Higgs (BEH) mechanism~\cite{Englert:1964et, Higgs:1964pj},
worked out a couple of years later for {\it relativistic} fields.

Nature illustrated SSB with ``squalid state'' physics,
which particle physics should be forever grateful.
But why should the Universe be permeated by a Higgs field since it began,
with a {\it v.e.v.} that gives SSB, is a mystery.

The remainder of this Review is organized as follows.
In Sec.~II we cover the Standard Model, in particular 
mass generation by coupling to the Higgs field,
as well as complex dynamics with three fermion generations.
In Sec.~III we put forth the Gell-Mann principle,
``{\it Everything not forbidden is compulsory.}'',
colloquially called ``the Totalitarian Principle'',
to discuss the possibility and implications for having a second Higgs doublet field,
establishing what we call the {\it General} two Higgs doublet model ({\it G}2HDM).
In Sec.~IV we address big issues related to {\it G}2HDM in the new Higgs/Flavor era,
namely:
 1)~baryogenesis confronts electron EDM; 
 2)~spectrum of extra Higgs bosons that are fit for search at the LHC; 
 3)~Nature's design in regards flavor. 
After a very brief discussion in Sec.~V, we offer our summary. 

We will keep this Review as nontechnical as possible.

\section{
Standard Model, Mass, and Higgs}

The only pivotal event so far at the LHC was in 2012:
observation~\cite{ATLAS:2012yve, CMS:2012qbp} of the Higgs boson, $h(125)$.
The use of the lower case $h$ would become clear in Sec.~III.
So why is it called the Higgs boson?
Because it was Higgs who showed~\cite{Higgs:1964pj}, after SSB of U(1),
a remnant particle would receive mass from its own {\it v.e.v.} 
This feature is retained in the Weinberg construction~\cite{Weinberg:1967tq} 
of the spontaneous breaking of the SU(2)$_L \times$U(1) symmetry 
by the {\it v.e.v.} of a complex doublet Higgs field.

Mass is the essence of matter, resulting in the clumpiness of 
the Sun, the Earth, etc. through gravitation.
So, the Higgs field is the source of mass in the Standard Model (SM)
of particle physics. Since all known massive fundamental particles
receive mass by their coupling to the {\it v.e.v.},
these particles would have definite couplings to the remnant scalar particle,
the Higgs boson $h$. Before the advent of the LHC,
the Higgs boson was whimsically dubbed ``the God particle'',
referring to its role as ``Origin of Mass''.
This all sounds simple, amazingly simple.

{\it \textbf{Dynamical Origin of Mass}.---} \
So, mass generation is dynamical in SM, by coupling to the universal condensate.
This is reasonable for the $W$ and $Z$ vector bosons,
the transmitter of the weak interaction, as it was the {\it raison d'\^ etre}.
What may be surprising is that Nature used practically the same token 
to generate fermion masses:
\begin{align}
M_V  = \frac{1}{2} gv
  \quad   \Longleftrightarrow   \quad 
m_f = \frac{1}{\sqrt 2}\lambda_f v.
\label{eq:m-propto-v}
\end{align}
%
The photon and the gluons remain massless
because they correspond to unbroken gauge symmetries.
Weinberg introduced~\cite{Weinberg:1967tq} the complex scalar double $\Phi$, 
\be
\Phi =
  \left[ \begin{array}{c}
        \phi^+ \\
        \phi^0
           \end{array} \right].
\label{eq:Phi}
\ee
The field develops a {\it v.e.v.} in the neutral component,
$\langle \phi^0\rangle = v$, and the SSB turns 
\be
  \left[ \begin{array}{c}
        \phi^+ \\
        \phi^0
           \end{array} \right]
  \ \longrightarrow \
  \left[ \begin{array}{c}
        G^+ \\
       v + h + iG^0
           \end{array} \right],
\label{eq:Phi-SSB}
\ee
where $G^\pm$ and $G^0$ are the (would-be) Goldstone bosons 
that become the longitudinal components of, i.e. ``eaten'' by,
the $W^\pm$ and $Z^0$ bosons to become massive.
But, as we will elucidate further in Sec.~III,
Weinberg also wrote down~\cite{Weinberg:1967tq} the other 
lepton-scalar interaction term, which gives 
the second, fermion part of Eq.~(\ref{eq:m-propto-v}).
Though usually called the Yukawa interaction, we will 
refer to $\lambda_f$ as the {\it Weinberg} coupling in remembrance.

The vector boson part of Eq.~(\ref{eq:m-propto-v}) reflects 
the prediction of the Glashow-Weinberg-Salam theory, 
which was affirmed by experiment, and the $W$ and $Z$ bosons
were directly observed at the S$p\bar p$S collider at CERN in 1983.
But turning further to the empirical nature of physics 
in the LHC era may be even more stunning.
First, the dynamical coupling of $g \simeq 2M_V/v$ was affirmed 
by the observation~\cite{ATLAS:2012yve, CMS:2012qbp} of 
$h \to WW^*$, $ZZ^*$ final states themselves,
and further confirmed in the observation of
Vector Boson Fusion (VBF) production of $VV \to h$~\cite{ATLAS:2016neq} 
via the $W$ and $Z$ bosons.
This may not be too surprising.
Second, during 2018--2020, a chain of measurements 
highlighted the prowess of the ATLAS and CMS experiments: 
both were able to measure the $tth$, $bbh$ and $\tau\tau h$
Weinberg couplings in 2018, and all agreed with SM expectations.

\begin{figure}[t]
\center
\includegraphics[width=0.4 \textwidth]{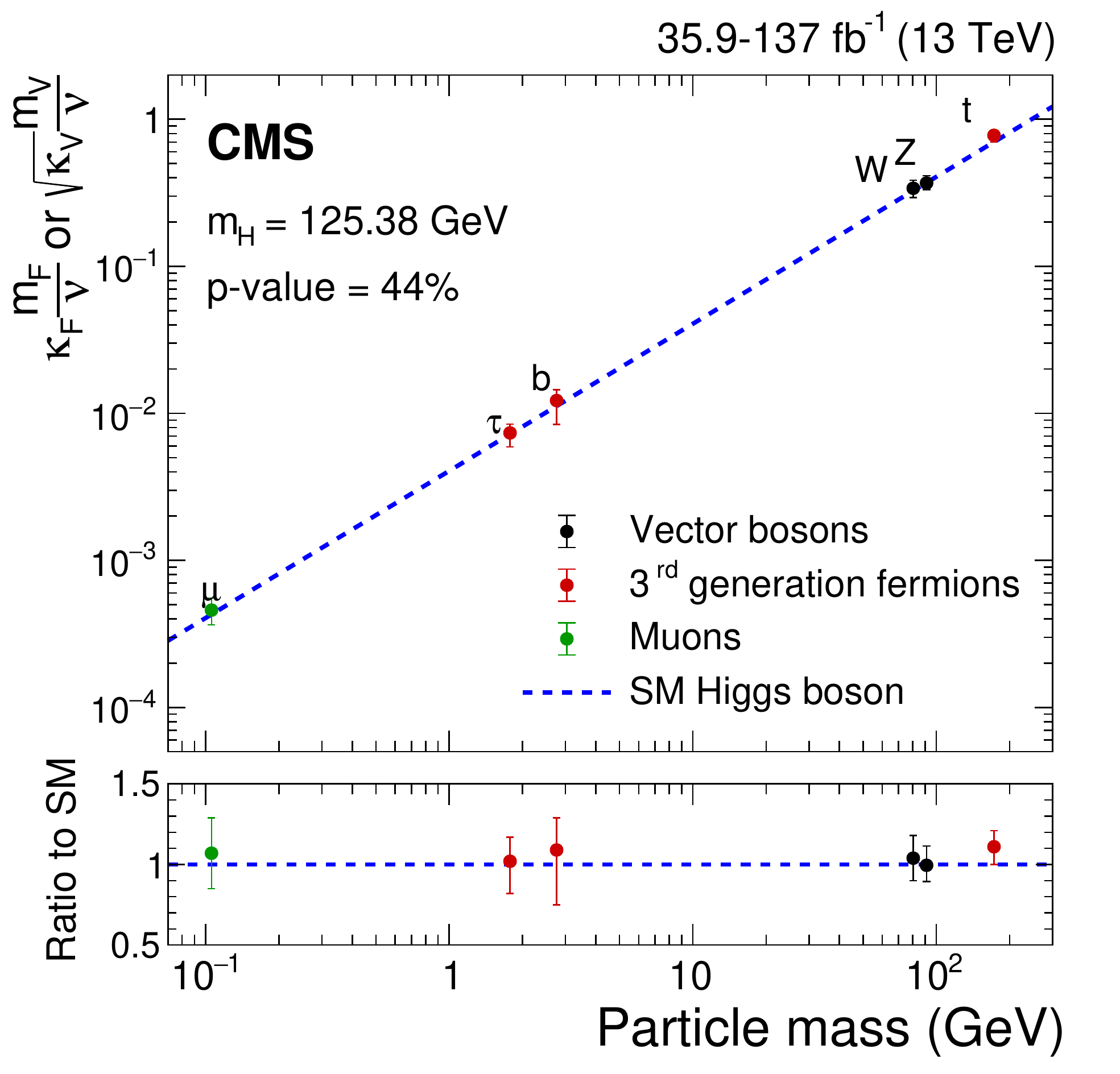}
\hskip0.1cm
\caption{
``Linear plot'' of $hVV$ and $hff$ couplings vs mass. 
 [Source: courtesy CMS Collaboration, from Ref.~\cite{CMS:2020xwi})
}
\label{fig:linear}
\end{figure}

Most impressive is the evidence for 
$h \to \mu\mu$ decay~\cite{CMS:2020xwi}, announced summer 2020,
where the measured couplings vs mass share the common slope
in the ``linear plot'' (see Fig.~\ref{fig:linear}), agreeing with  
\begin{align}
 \frac{1}{2} g & = M_V / v, \\
\frac{1}{\sqrt 2} \lambda_f & = m_f / v.
\label{eq:coup-propto-v}
\end{align}
That this covers  6 particles, and from $\lambda_t \simeq 1$ 
to $\lambda_\mu \simeq 0.0006$, spans over more than 
3 orders in coupling strength (or mass), is truly eye-popping. 
The {\it dynamical} nature of mass generation is now empirical.

If Nature chooses this very simple scheme for generating particle masses, 
the way she generates SSB, i.e. nonvanishing {\it v.e.v.} is also deceivingly simple.
One writes the very simple Higgs potential, 
\begin{align}
 V(\Phi)  =  \mu^2|\Phi|^2 + \lambda |\Phi|^4,
\label{eq:pot}
\end{align}
first for a complex scalar $\Phi$, but works also for a complex doublet.
Demand now $\mu^2 < 0$, one gets
(the bottom of) a ``wine bottle'' shape,\footnote{
 Nomenclature preferred by Peter Higgs.
}   
with minima at $|\langle \phi^0\rangle|^2 = v^2 = |\mu^2|/\lambda$.
The purist would say we do not really know the shape of the Higgs potential,
which is true. But Occam's razor would favor the simplicity of Eq.~(\ref{eq:pot}).

{\it \textbf{Chiral Fermions and Complex Couplings}.---} \
One really curious aspect for fermion mass generation in SM is that,
the weak interaction is {\it left-handed}. Had the weak interaction been
vector-like as in the Dirac theory of the electron (QED), or like QCD,
then gauge dynamics would dictate all coupling constants be real!

Now take the weak coupling, 
\begin{align}
i\,gV_{ij}\, \bar u_i \gamma_\mu Ld_j W_\mu, 
\label{eq:udW}
\end{align}
where $V$ is the CKM matrix, and only left-handed quarks participate. 
Replacing $W_\mu$ by $k_\mu/M_W$ and imposing on-shell condition for external quarks $u_i$ and $d_j$, 
after simple manipulations~\cite{Hou:2012az}, one finds
\begin{align}
i\,V_{ij}\, \bar u_i \left(\frac{m_i}{v} L - \frac{m_j}{v} R\right) d_j G, 
\label{eq:udG}
\end{align}
where the gauge coupling $g$ cancels out the $g$ in $M_W = gv/2$. 
Note that $k_\mu k_\nu/M_W^2$ propagates 
the would-be Goldstone boson, we find that 
the would-be (i.e. eaten) Goldstone boson $G$ couples with Weinberg coupling,
and hides behind, or is veiled by, the vector gauge coupling.
We have thus reverse engineered the Weinberg coupling
$\lambda_f$ using empirical knowledge;
 it has more to do with the left-handed doublet, right-handed singlet 
nature of fermions under the SU(2)$_L$ gauge group, 
rather than the Higgs representation used for SSB.

We will elucidate in Sec.~III why Weinberg couplings, 
unlike gauge couplings, are intrinsically complex.
For now, recall the textbook example of the $3\times 3$ matrix $V$.
Motivated by the observation~\cite{Christenson:1964fg} 
of CPV, Kobayashi and Maskawa showed~\cite{Kobayashi:1973fv} 
there is a unique, irremovable phase.
In Eq.~(\ref{eq:udW}), one may be confused to think that 
the KM phase is associated with the gauge sector,
but it arises from diagonalization of quark masses,
which arise dynamically from Weinberg couplings.
The matrix $V$ is the difference of the $u$- and $d$-type
left handed transformations of the respective bi-unitary transforms.

The unitarity of the CKM matrix has been experimentally 
affirmed at the per mille level~\cite{Chang:2017wpl},
and usually parametrized in the Wolfenstein form~\cite{Wolfenstein:1983yz},
\be
 V 
 \simeq
    \left(  \begin{array}{ccc}
    1 - \lambda^2/2               &     \lambda     &  A\lambda^3(\rho - i\,\eta) \\
      - \lambda                   & 1 - \lambda^2/2 &  A\lambda^2 \\
    A\lambda^3(1 - \rho - i\, \eta) &   -A\lambda^2   &  1
   \end{array} \right),
 \label{eq:V_Wolf}
\ee
where the CPV phase is placed in $V_{ub}$ and $V_{td}$,
$\lambda \equiv V_{us} \cong 0.22$ is the Cabibbo angle, and
\be
A\lambda^2 \simeq 0.041, \ 
A\lambda^3\sqrt{\rho^2+\eta^2}  \sim 0.0036, \
 \label{eq:VcbVub}
\ee
are the strengths of $V_{cb}$ and $V_{ub}$, respectively.
Agonizingly, the KM phase in the $V$ matrix 
--- confirmed by BaBar~\cite{BaBar:2001pki} and Belle~\cite{Belle:2001zzw} ---
is by far (roughly by $10^{-10}$) insufficient~\cite{Hou:2008xd} 
for the CPV condition of Sakharov to generate BAU.
Furthermore, SM cannot provide the other condition, namely out of equilibrium,
and electroweak phase transition in SM is a crossover.
So, as the B factories were still even under construction, 
the experimenters had more than one reason to 
prefer {\it ruling against} the KM model. 
But Nature's design was confirmed instead: 
all laboratory measurements of CPV so far 
can be explained by the KM phase.

\section{Gell-Mann Principle and 
{\it G}2HDM
}

{\it G}2HDM in the section title stands for {\it General} two Higgs doublet model.
Having found empirically a Higgs boson weak doublet,
even though three components appear as  would-be Goldstone bosons,
it seems totally reasonable to have a second doublet in Nature.
But we wish to argue how this second doublet should ``behave'',
i.e. what its existence would bring forth.

{\it \textbf{Three Types of 2HDM}.---} \
Our first work (preceded slightly by Ref.~\cite{Grinstein:1987vj}) 
involving 2HDM was on the charged Higgs ($H^+$) effect 
on $b \to s\gamma$~\cite{Hou:1987kf}, which adopted 
the ``Natural Flavor Conservation'' (NFC) condition~\cite{Glashow:1976nt}
 of Glashow and Weinberg.
Basically, in Model\;I, both $u$- and $d$-type quarks receive mass
from just one scalar doublet, while in Model\;II, they receive mass from separate doublets.
Because it automatically appears with supersymmetry (SUSY), 
2HDM\;II is particularly popular. 
We found that in 2HDM\;II, the $\tan\beta$ factor associated with $d$-type quarks
cancels out the $\cot\beta$ factor associated with $u$-type quarks, leaving an
$H^+$ correction term that is always constructive w.r.t. the SM effect.
As much as $b \to s\gamma$ developed into its own cottage industry,
after it was observed by CLEO~\cite{CLEO:1994veu},
the reason that it provides the most stringent bound on $m_{H^+}$ so far
is rooted in this short-distance effect.
The effect in 2HDM\;I, however, is destructive, and less studied.

Our second well-known 2HDM work is~\cite{Hou:1992sy} 
$B \to \tau\bar\nu$ decay.
Having proposed~\cite{Grzadkowski:1991kb} two possible solutions to 
the then ``$B$ meson semileptonic branching ratio puzzle",
one of which is enhanced $b \to c\tau\nu$ to ${\cal O}(10\%)$,
only to see it ruled out by ALEPH~\cite{ALEPH:1992zwu}.
By curiosity, we checked $B \to \tau\nu$, as it had not been touched upon.
We found the $H^+$ effect in 2HDM\;II to be destructive against SM,
amounting to a multiplicative factor over the SM expectation
that is {\it independent} of hadronic parameters. 
Once evidence emerged~\cite{Belle:2006but} at Belle, 
the process became the second most powerful bound on the $H^+$ boson.

Our third better known 2HDM work went beyond Models I \& II.
Although the NFC condition of Glashow and Weinberg sounded convincing, and
going hand in hand with SUSY made 2HDM~II quite popular,
it was pointed out a decade later that, to demand
each type of quarks receive mass from 
{\it just one} scalar doublet when two are present, may be overkill,
i.e. maybe not needed.
One motivation is the Fritzsch ansatz~\cite{Fritzsch:1977za},
%
\be
  {\boldmath m}_d = \left[ \begin{array}{cc}
       0 & \sqrt{m_d m_s} \\
        \sqrt{m_d m_s} & m_s
           \end{array} \right].
\label{eq:Fritzsch}
\ee
which connects the Cabibbo angle and $m_d/m_s$ 
mass ratio: $\sin\theta_C \simeq \sqrt{{m_d}/{m_s}}$.
Perhaps further stimulated by the ARGUS observation~\cite{ARGUS:1987xtv} 
of finite $B^0$--$\overline B^0$ mixing,
Cheng and Sher extended~\cite{Cheng:1987rs} the Fritzsch ansatz 
to 3 generations, with the trickling-off nature of $\sqrt{m_im_j}$ 
as one goes off-diagonal, stressing that this provides some control of
flavor changing neutral Higgs (FCNH) couplings, reducing the need for NFC.

A few years later, Sher proposed~\cite{Sher:1991yb} 
the search for $b' \to bh$, where $h$ is a Higgs boson of the 2HDM,
and $b'$ the 4th generation (4G) $d$-type quark. 
We learned of this paper because three out of its five references referred to 
$b' \to bH$ decay~\cite{Hou:1988yu, Haeri:1988jt, Hou:1990wz}, 
which are loop-induced through the $t'$ quark, an SM albeit 4G effect.
We noticed that the extra assumption of 4G was not necessary,
and immediately pointed out~\cite{Hou:1991un} the process to look out for is 
$t \to ch$, or $h \to t\bar c$ if the generic exotic Higgs boson with FCNH $tch$ coupling
turns out to be heavier than the top quark.
We further stressed that it is the fermion {\it mass and mixing hierarchies},
rather than the specific form of the Cheng-Sher ansatz, that
allows one to  drop the NFC condition. 
Since this is neither Model I nor II, the terminology of 2HDM~III was suggested.

{\it G}2HDM is basically 2HDM~III, where further developments
would be elucidated in the next section.

{\it \textbf{Gell-Mann Principle and a Second Higgs}.---} \
Since 1992, we on-and-off worked on 2HDM~III,
e.g. elucidating~\cite{Chang:1993kw} the two-loop (Barr-Zee/Bjorken-Weinberg) 
mechanism for $\mu \to e\gamma$, facilitated by a top loop,
which also applies to $\tau \to \mu\gamma$~\cite{Harnik:2012pb}.

In revisiting $\mu \to e\gamma$ and other muon Flavor Violation 
($\mu$FV) processes recently~\cite{Hou:2020itz}, 
perhaps influenced by our living experience in Germany,
one day the verse ``what is not forbidden is allowed'' came to mind.
Searching the web, we found a Wikipedia entry
 ``Everything which is not forbidden is allowed''. 
But, surprisingly, the second entry of the search 
was the {\it Totalitarian principle}: 
\be
{``Everything\ not\ forbidden\ is\ compulsory."}
\label{pr: G-M}
\ee
\noindent which was traced to Gell-Mann~\cite{Gell-Mann:1956iqa},
with the words ``Anything that is not compulsory is forbidden.''
used both ways. Gell-Mann referred to the state of affairs 
in ``a perfect totalitarian state'', but in the context of strong interactions: 
``any process not forbidden by a conservation law does 
take place with appreciable probability.''
Let us concern not with logical equivalence~\cite{Kragh1907} here,
but we shall call ``Eq.~(\ref{pr: G-M})'' the Gell-Mann Principle.

Since we now know firmly one Higgs doublet exists, and we know of
no reason to forbid a second Higgs, the simple application of 
the Gell-Mann Principle dictates that 
a second scalar doublet, $\Phi'$, {\it must} exist (G-M:1).
As we will see in the next section, it is convenient to have $\Phi$ generating
the {\it v.e.v.}, hence $\langle \phi'^0\rangle = 0$ for $\Phi'$,
which does not contribute to {\it v.e.v.} Thus,
\be
  \Phi' = 
  \left[ \begin{array}{c}
        \phi'^+ \\
        \phi'^0
           \end{array} \right]
  \ \longrightarrow \
  \left[ \begin{array}{c}
        H^+ \\
       H + iA
           \end{array} \right],
\label{eq:Phi'}
\ee
after SSB is induced by $\langle \phi^0\rangle = v$,
where $H$ and $A$ are the exotic neutral scalar and pseudoscalar.
We will discuss the Higgs potential in the next section.

One may immediately ask, ``why not three scalar doublets?''
While legitimate, let us say that this 2HDM, in particular 
the {\it G}2HDM that we will discuss, has enough new parameters, 
such that, short of model building, which we do not pursue, 
it is prudent to proceed with one extra doublet at a time (principle of simplicity).

The emergence of $H$, $A$ and $H^+$ exotic Higgs bosons in 2HDM is 
quite familiar, with the natural questions: 
``Where are they?'', and ``What do they do?''
We therefore return to Weinberg's~\cite{Weinberg:1967tq} 
original {\it Model of Leptons},
and the NFC work with Glashow~\cite{Glashow:1976nt},
with an eye towards upholding the Gell-Mann Principle, Eq.~(\ref{pr: G-M}).

{\it \textbf{Weinberg Coupling and Gell-Mann Principle}.---} \
To spontaneously break the SU(2)$_L \times$U(1) symmetry~\cite{Glashow:1961tr} 
down to U$_Q$(1), Weinberg introduced a weak scalar doublet field 
$\Phi$, Eq.~(\ref{eq:Phi}), and demanded $\langle \phi^0\rangle = v$, 
as we have already seen. Charge symmetry, corresponding to QED, is not broken.
However, the essence is that the left-handed leptons transform as a weak doublet
\be
  \ell_L = 
  \left[ \begin{array}{c}
          \nu_L \\
            e_L
           \end{array} \right],
\label{eq:ell}
\ee
while the right-handed $e_R$ transforms as a singlet.
Weinberg followed what one is taught in field theory: keep all terms allowed
by symmetry. Hence, he wrote down
\be
  \lambda_e \bar\ell_L \Phi e_R,
\label{eq:ell_Phi_eR}
\ee
which is dimension-4 and gauge singlet, 
hence should be included in the Lagrangian.
Upon SSB, the electron gains mass as we saw in Sec.~II.
So, $\lambda_e$ is the {\it Weinberg coupling} for the electron, 
responsible for $m_e$ generation.
Subsequently, this formulation was generalized to three families
for all types of charged fermions, 
namely
\be
  \sum_{i,j} \left(\bar\ell_{iL} \Lambda_{ij}^e e_{jR}
         +    \bar q_{iL} \Lambda_{ij}^u u_{jR}
         +    \bar q_{iL} \Lambda_{ij}^d d_{jR}  \right) \Phi
         +    h.c.,
\label{eq:WeinCoup}
\ee
where $q_{iL} \equiv [u_{iL}, d_{iL}]$, 
{\boldmath $\Lambda$}$^f$ are $3 \times 3$ 
Weinberg coupling matrices for $f = e, d, u$.
A point to stress is that, because of the chiral nature
of the weak interaction, the {\boldmath $\Lambda$}$^f$ matrices are {\it complex}. 
The usual program, then, is to diagonalize {\boldmath $\Lambda$}$^f$
by biunitary transforms, with real eigenvalues $\lambda_f$,
where $f$ now covers all charged fermions,
and $V$ is the difference between $U_L$ and $D_L$,
the left-handed transforms for $u$- and $d$-type quarks, respectively.
The observation of Kobayashi and Maskawa is that,
using all phase freedoms of quark fields, 
one irremovable phase remains~\cite{Kobayashi:1973fv} in $V$, 
as seen in Eq.~(\ref{eq:V_Wolf}).

Up to this point, on one hand one may find it deficient
that the nine $\lambda_f$'s span 6 orders of magnitude in strength,
and wonder what Nature has up her sleeves (which is part of the flavor problem). 
On the other hand, let us point out that the statement  above,
\be
``keep\  all\ terms\ allowed\ by\ symmetry" \nn
\ee
of the Lagrangian, 
is nothing but the Gell-Mann Principle! 
As this preceded G-M:1, the application of Gell-Mann Principle
to dictate the existence of a second scalar doublet,
the mandatory existence of Weinberg interaction can be classified as G-M:0.

So, introducing the complex scalar doublet for sake of  spontaneous
breaking of SU(2)$_L \; \times \;$U(1) to U$_Q$(1)
is necessarily accompanied by the Weinberg interaction. 
Thus, Eq.~(\ref{eq:m-propto-v}), or vector boson and fermion
mass generation, are indeed achieved in one stroke.

{\it \textbf{To NFC or Not To NFC}.---} \
We can now turn to the NFC condition~\cite{Glashow:1976nt} of Glashow and Weinberg.

Having invented the GIM mechanism~\cite{Glashow:1970gm} to explain 
the very strong suppression of e.g. $K_L \to \mu^+\mu^-$ vs $K^+ \to \mu^+\nu$,
Glashow certainly had legitimate worries about FCNH when 
there is a second Higgs doublet.
To wit, introducing the second doublet $\Phi'$, there should be
a second Weinberg interaction term,
\be
  \sum_{i,j} \left(\bar\ell_{iL}P_{ij}^e e_{jR}
         +    \bar q_{iL} P_{ij}^u u_{jR}
         +    \bar q_{iL} P_{ij}^d d_{jR}  \right) \Phi'
         +    h.c.,
\label{eq:WeinCoup'}
\ee
where $P$ stands for the Capital Greek letter for $\rho$.
Regardless of how we assign {\it v.e.v.} generation, 
the {\boldmath $\Lambda$}$^f$ and {\boldmath $P$}$^f$ matrices 
{\it cannot} be simultaneously diagonalized,
and there would be FCNH couplings in general.
This is the reason behind the Glashow-Weinberg NFC condition,
which declares that for each fermion mass matrix, there can be 
only one Weinberg (usually called Yukawa) coupling matrix, 
i.e. absence of a second Weinberg matrix.
Then, the fermion mass matrices are necessarily simultaneously diagonalized
with the Weinberg matrices, and there are no FCNH couplings.
{\it All fear is gone!}

But this is no less an {\it Edict}, as given by princes, hence
authoritarian, and not in the scholarly tradition.
Now, NFC is readily enforced by introducing a $Z_2$ {\it symmetry},
giving rise to 2HDM\;I \& II, which correspond to linking
the two Higgs doublets to $u$- and $d$-type quarks.
Invoking a symmetry elevates the standing of NFC superficially 
in the context of the Gell-Mann Principle.
It was also very {\it convenient} that it coincided with 
the outcome of SUSY, making 2HDM\;II popular.
Alas, if SUSY had emerged quickly with the advent of the LHC,
this picture would have gained credence. 
But we have not discovered {\it anything} beyond 
the $h(125)$, which by all counts resembles the Higgs boson of SM!
Indeed, with the mindset of SUSY, it is easy to think of
the extra doublet as decoupled~\cite{Kane:2018oax}, 
i.e. multi-TeV and out of reach.

But, should we give up $H$, $A$ and $H^+$ search?
From the experimental point of view, the answer is clearly an emphatic ``{\it No!}\,''
Let LHC run its course, and finish what it is designed to do.
We argue that NFC should be viewed the way it really is: {\it ad hoc}.
It is NFC, and the familiar $Z_2$ symmetry, that should be dropped.
Since the LHC has not revealed any new symmetry,
we call the situation -{\it Dyssymmetry}\,-, playing on the word dissymmetry.
After all, the Gell-Mann Principle should reign higher than
the edict of NFC by Glashow and Weinberg.
To put it differently, to remove {\it all} extra Weinberg matrices in 2HDM
for fear of FCNH is itself unseemly.

{\it \textbf{{\it G}2HDM and the Gell-Mann Principle}.---} \
If one drops the $Z_2$ symmetry for implementing NFC,
and use G-M:1 to dictate the existence of a second scalar doublet
after completion of the first doublet, iterating the Gell-Mann Principle
a second time, or G-M:2, would dictate that the {\boldmath $P$}$^f$ 
matrices should exist for all $f = e,\,d,\,u$. 
Undergoing the same biunitary transform that 
diagonalizes the {\boldmath $\Lambda$}$^f$ matrices,
Eq.~(\ref{eq:WeinCoup'}) becomes
\be
  \sum_{i,j} \left(\bar\ell_{iL}\rho_{ij}^e e_{jR}
         +    \bar q_{iL} \rho_{ij}^u u_{jR}
         +    \bar q_{iL} \rho_{ij}^d d_{jR}  \right) \Phi'
         +    h.c.,
\label{eq:Wein-coup'}
\ee
where {\boldmath $\rho$}$^f$ matrices are in general not diagonal
and complex.
The Cheng-Sher ansatz~\cite{Cheng:1987rs} of $\rho_{ij} \sim \sqrt{m_i m_j}$ 
paved the way, but in generalizing to 2HDM~III~\cite{Hou:1991un}, 
we stressed that it is the fermion mass-mixing hierarchies that seem to
tame what Glashow had feared for the worse.

What did Glashow and Weinberg  not know, in 1977?
They were well aware of the fermion mass hierarchies,
%
\begin{align}
    & m_e  \ \ll\  m_\mu  \ \ll\  m_\tau, \nn \\
    & m_d  \ \ll\  m_s  \ \ll\  m_b, \label{eq:mass-hier} \\
    & m_u  \ \ll\  m_c  \ \ll\  m_t, \nn 
\end{align}
as all masses, sans the top, were known in 1977.
But then {\it nobody anticipated how large the ratio}
\be
 {m_t/m_b \gg 1},
\label{eq:toverb}
\ee
would be. The first hint came only with 
the ARGUS discovery~\cite{ARGUS:1987xtv} 
of large $B_d$--$\overline B_d$ mixing in 1987.
Given that $m_b \ll v \approx 246$~GeV,
who would have thought $\lambda_t \simeq 1$!!

If $m_t/m_b \gg 1$ was not anticipated, what were again purely 
empirical are the discoveries, circa 1983, of the long $b$ lifetime, 
and the absence of $b \to u$ vs $b \to c$ transitions. 
Sniffing these out, the experimental discoveries 
--- divinations of Nature --- inspired Wolfenstein~\cite{Wolfenstein:1983yz} 
to write down his seminal parameterization, Eq.~(\ref{eq:V_Wolf}).
Another thing unaware to Glashow and Weinberg --- perhaps they would not care ---
at time of their writing Ref.~\cite{Glashow:1976nt},
is the Fritzsch ansatz~\cite{Fritzsch:1977za}, Eq.~(\ref{eq:Fritzsch}),
which curiously related the $m_d/m_s$ ratio to the Cabibbo angle.
But we should stress that the quark mixing hierarchy
\be
{|V_{ub}|^2 \ll |V_{cb}|^2 \ll |V_{us}|^2} \ll  |V_{tb}|^2 \cong 1,
 \label{eq:mixing-hier}
\ee
is staggering, with $|V_{us}|^2 \simeq 0.05$, $|V_{cb}|^2 \sim 10^{-3}$,
 $|V_{ub}|^2 \sim 10^{-5}$.
It is the combination of Eqs.~(\ref{eq:mass-hier})--(\ref{eq:mixing-hier}),
{\it all emergent}, that Nature seems to employ to {\it hide}
the effects of the second scalar doublet from us, so far.
This is curiously against the spirit of the Gell-Mann Principle, 
but we have to admit that Nature can have her ways.

{\it \textbf{FCNH: an Experimental Question}.---} \
A final point to make, before we move on to explore {\it G}2HDM
and its consequences in the next section, is returning to 
the empirical nature of physics, yet again.

At the time it was proposed~\cite{Hou:1991un}, 
be it $t \to ch^0$ or $h^0 \to t\bar c$, where $h^0$ is some neutral scalar boson,
the top quark was not yet observed, and neither was any fundamental scalar boson.
It was therefore a theoretical conjecture inspired by 
the Cheng-Sher ansatz~\cite{Cheng:1987rs}, depending on
whether $t$, or ${h^0}$, is the heavier one.
We did extend beyond the Cheng-Sher ansatz and called it 2HDM\;III,
i.e. neither 2HDM\;I nor II, but one where there is no $Z_2$ symmetry
to enforce NFC. 
This {\it general} (not {\it ad hoc}) 2HDM that possesses extra Weinberg couplings,
which includes FCNH, we now call {\it G}2HDM.

What happened after observation of $h(125)$ in 2012?
If one looks up Particle Data Group~\cite{PDG},
one would find entries for bounds\footnote{
 We continue to use $h$ for the 125~GeV boson. 
 Experiments still use $H$, as nothing else has been discovered so far.
} 
on $t \to ch$ from ATLAS~\cite{ATLAS:2014lfm} 
and CMS~\cite{CMS:2014jkv}, dating back to 2014.
But, in particular ATLAS had already shown results
for $t \to qh$ with $h \to \gamma\gamma$ at the EPS-HEP conference
held 2013 in Stockholm.
The point is, with $h(125)$ found lighter than top, 
there is no way to stop the experimentalists, young or not
(e.g. Daniel Fournier of ATLAS) from $t \to ch$ search --- {\it because they can}.
Suppose someone resorts to authority and tells the valiant worker that
Glashow and Weinberg forbade it since 1977,
the first response would be shock: ``{\it What\,!?}\,'' 
But after a brief pause, the person would brighten up and say 
``Splendid, so much the better. {\it Can Not Lose\,!\,}"
Because the bottom line would be: {\it It is a PDG Entry.}
The current bound~\cite{CMS:2021bdg} at 95\% C.L. is 
\be 
{\cal B}(t \to ch) < 7.3 \times 10^{-4}. \quad {\rm (CMS\, 2021)}
\label{eq:tch}
\ee

A second case is $h \to \tau\mu$.
Advocated by Ref.~\cite{Harnik:2012pb}, in fact
CMS saw a hint~\cite{CMS:2015qee} at 2.4$\sigma$ with 
branching fraction $\sim 1\%$ in 2015,
but disappeared when more data was added.
The current bound~\cite{CMS:2021rsq} at 95\% C.L. is 
\be 
{\cal B}(h \to \tau\mu) < 1.5 \times 10^{-3}. \quad {\rm (CMS\, 2021)}
\label{eq:htamu}
\ee

So, FCNH is an experimental question. 
And by generalization, it should be extended to 
{\it all} extra Weinberg couplings $\rho_{ij}^f$,
for $i,\, j = $1--3 and $f = u,\,d,\,e$.

The extra Weinberg couplings in {\it G}2HDM are~\cite{Davidson:2005cw,Hou:2019mve}
\begin{align}
 & - \bar{\nu}_i\rho^\ell_{ij} R \, \ell_j H^+
    - \bar{u}_i\left[(V\rho^d)_{ij} R - (\rho^{u\dagger}V)_{ij} L\right]d_j H^+ \notag\\
 & - \frac{1}{\sqrt{2}} \sum_{f = \ell,}^{u, d} \bar f_{i}\Big[
 \Big.\big(\lambda^f_i \delta_{ij} c_\gamma + \rho^f_{ij} s_\gamma\big)H
 - i\,{\rm sgn}(Q_f) \rho^f_{ij} A \Big. \notag\\
 &\quad\quad\quad\quad\quad
  - \big(\lambda^f_i \delta_{ij} s_\gamma - \rho^f_{ij} c_\gamma\big) h \Big]  R\, f_{j}
 +{h.c.},
\label{eq:Wein}
\end{align}
where $i$, $j$ are summed over generations, 
$L, R$ are projection operators, and $V$ is the CKM matrix, 
with lepton matrix taken as unity due to vanishing neutrino masses.
%
%
With $h$ identified as the observed SM-like Higgs boson,
one can read off Eq.~(\ref{eq:Wein}) and see that
$c_\gamma$ is the $h$-$H$ mixing angle
between the two $CP$-even scalars.
As we shall see in the next section, the nonobservation of $t \to ch$ 
and $h \to \tau\mu$ so far may have another surprise of Nature behind it:
small $c_\gamma$, that $h$, the SM sector, and $H$, 
the exotic sector, do not seem to mix much.

\section{Big Issues \& New Higgs/Flavor Era}

\vskip0.2cm

{\it
\quad\quad\quad\quad\quad O Lord, our\ Lord, 

\quad\quad\quad\quad\quad \quad How Majestic is Thy Name 

\quad\quad\quad\quad\quad \quad\quad\quad in all the {\cal Earth}, 

\quad\quad\quad\quad\quad \quad Who have set Thy Splendor 

\quad\quad\quad\quad\quad \quad\quad\quad  above the {\cal Heavens}\,!
}

\vskip-0.5cm 
%
\subsection{Heaven and Earth}

By {\it Heaven} we refer to baryogenesis. 
In particular, electroweak baryogenesis (EWBG) is preferred 
by down to {\it Earth} physicists, such as at the LHC, or
those working on exquisite low energy precision experiments (LEPEs).
Take any pair of the three, however, there is tension.

Baryogenesis calls for additional CPV that is beyond SM (BSM), 
i.e. beyond the Kobayashi-Maskawa phase.
EWBG, which further requires a strongly first order electroweak phase transition
 (1$^{\rm st}$EWPT), is attractive because it is more testable. 
But hereby lies the tension: EWBG calls for large BSM CPV,
but we have not seen any BSM physics at the LHC so far!

EWBG also easily runs into tension with LEPEs,
where a recent flagship example is the ACME experiment
aimed at electron electric dipole moment (eEDM).
The current ACME bound~\cite{ACME:2018yjb} on eEDM is,
\be
|d_e| < 1.1 \times 10^{-29}\; e\, {\rm cm}, \quad ({\rm ACME}\, 2018)
\label{eq:ACME18}
\ee
at 90\% C.L. As EWBG demands rather large BSM CPV, as well as
new particles with sufficient dynamics to instigate 1$^{\rm st}$EWPT,
it is usually not trivial to evade the probing eye of LEPEs such as ACME.
As we shall see, EWBG can be achieved~\cite{Fuyuto:2017ewj}
in {\it G}2HDM, but its initial prediction for eEDM was 
ruled out by ACME\,2018, Eq.~(\ref{eq:ACME18}), within a year!
Fortunately, Nature seems to provide a mechanism~\cite{Fuyuto:2019svr} 
to render $d_e$ small.

Finally, the tension between LHC experiments and LEPEs is, plainly,
 {\it asymmetric competition}.
While we have not observed any BSM physics at the LHC,
the ``table-top'' LEPEs (which ACME truly is),
with its claim to be able to probe higher scales,
causes further angst for the behemoth LHC experiments.

{\it \textbf{EWBG in {\it G}2HDM}.---} \
The CMS hint~\cite{CMS:2015qee} for $h \to \tau\mu$
caused some stir and hope in the HEP circle.
In Fall 2016, there was a talk on EWBG that combined 
$h \to \tau\mu$ with $\tau \to \mu\gamma$,
employing large $\rho_{\tau\mu}$ or $\rho_{\tau\tau}$
 (vs $\lambda_\tau \simeq 0.01$) as 
CPV source~\cite{Chiang:2016vgf} for baryogenesis.
After the seminar, we spoke with the speaker and commented: 
``This is too intricate for it to work in Nature.'', remarking further, 
``However, $\rho_{tt}$ is naturally ${\cal O}(1)$ and complex. 
I would bet that it can provide EWBG.''

The context of the talk was basically {\it G}2HDM (i.e. 2HDM~III~\cite{Hou:1991un}),
and we learned from the talk that the extra Higgs bosons in 2HDM 
could in principle provide 1$^{\rm st}$EWPT. 
And finally, we started to pay attention to the CPV 
phases that {\it G}2HDM provide, and set off to work. 
The results turned out better than anticipated.

As seen from Fig.~\ref{fig:linear}, with the Weinberg coupling 
$\lambda_t \cong 1$ in SM, it is plausible that
$\rho_{tt}$ and $\rho_{tc}$ are ${\cal O}(\lambda_t)$ 
hence\footnote{
The other element, $\rho_{ct}$, 
has to be relatively small~\cite{Altunkaynak:2015twa}
as it enters e.g. $b \to s\gamma$ with chiral enhancement.
} 
order 1. This is the reason behind our comment on the leading effect for EWBG.
We thereby showed~\cite{Fuyuto:2017ewj} that
\be
  \lambda_t\,{\rm Im}\, \rho_{tt},
\label{eq:CPV-BAU}
\ee
provides a robust driver for BAU. 

\begin{figure}[t]
\center
\includegraphics[width=0.34 \textwidth]{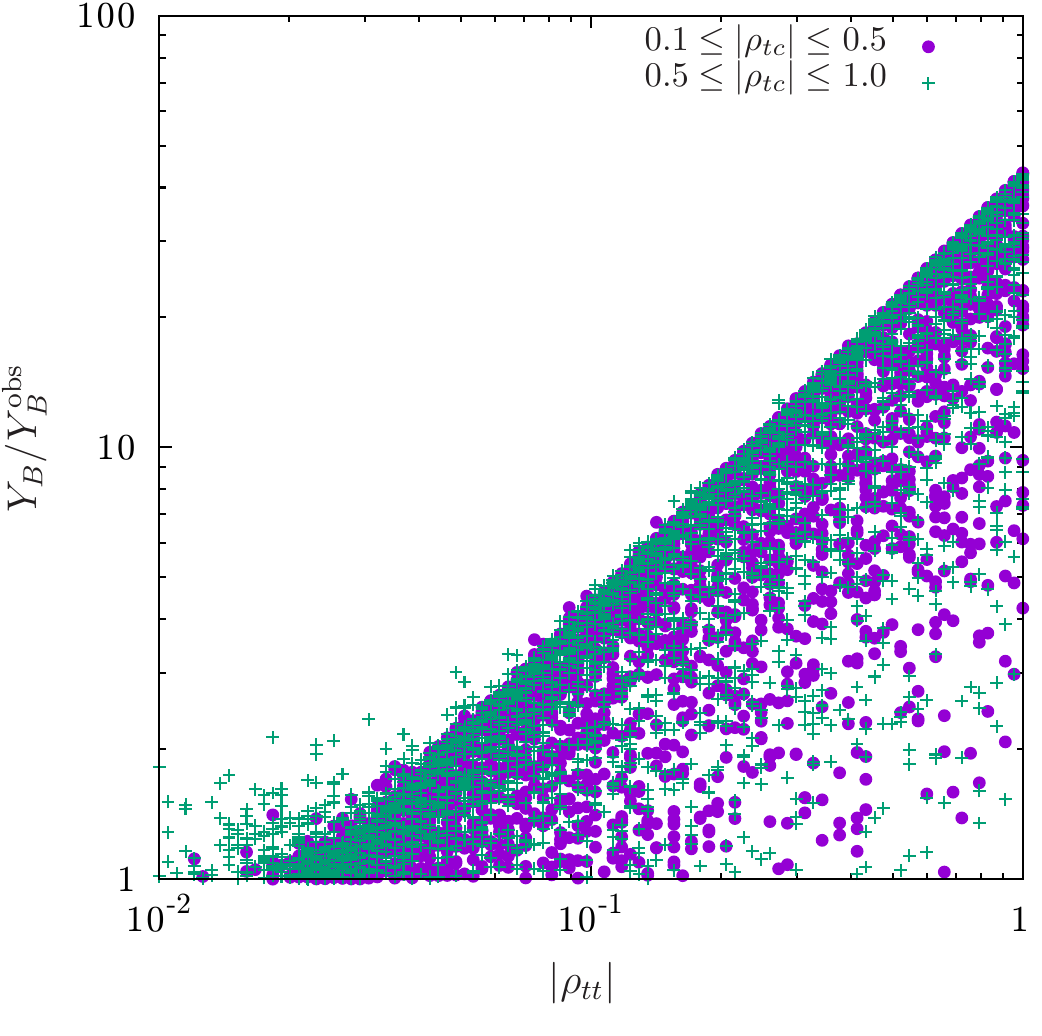}
\caption{
$Y_B/Y_B^{\rm obs}$ vs $|\rho_{tt}|$. 
The phases $\arg \rho_{tt}$ and $\arg \rho_{tc}$ are scanned over,
with various parameters randomly chosen, and 
taking $m_H = m_A = m_{H^+} =$ 500~GeV.
The purple bullets (green crosses) are for $0.1 \leq |\rho_{tc}| \leq 0.5$
($0.5 \leq |\rho_{tc}| \leq 1.0$).
Plot taken from Ref.~\cite{Fuyuto:2017ewj},
which we refer to for further details.
}
\label{fig:EWBG}
\end{figure}

This is illustrated in Fig.~\ref{fig:EWBG}.
The baryon over entropy, $Y_B = n_B/s$, 
is plotted~\cite{Fuyuto:2017ewj} in units of the observed~\cite{Planck:2013pxb} 
\be
 Y_B^{\rm obs} \simeq 8.59 \times 10^{-11}, \quad {\rm (Planck\,2013)}
\label{eq:Planck2013}
\ee
vs the strength of $\rho_{tt}$,
where we scan over the phases of $\rho_{tt}$ and $\rho_{tc}$.
To further discern the cause and effect, we plot the scatter points
for $0.1 \leq |\rho_{tc}| \leq 0.5$ ($0.5 \leq |\rho_{tc}| \leq 1.0$)
in purple bullets (green crosses). 
For the bulk of the plot, the two sets can be barely distinguished,
so $\rho_{tt}$ is the driver for BAU and appears quite robust. 
But for $|\rho_{tt}| \lesssim 0.07$ or so, the main effect seems to peter out,
and we find that $1 < Y_B/Y_B^{\rm obs} \lesssim 2$-3
to be populated more by $0.5 \leq |\rho_{tc}| \leq 1.0$.
This indicates that, when $|\rho_{tt}| \lesssim 0.05$ and becomes ineffective,
$\rho_{tc}$ at ${\cal O}(1)$ provides a second, 
backup~\cite{Fuyuto:2017ewj} mechanism for BAU.
This was also illustrated heuristically with numerical 
$2 \times 2$ examples for the 2-3 sector of the {\boldmath $\rho$}$^u$ matrix.

\begin{figure}[b]
\center
\includegraphics[width=0.2 \textwidth]{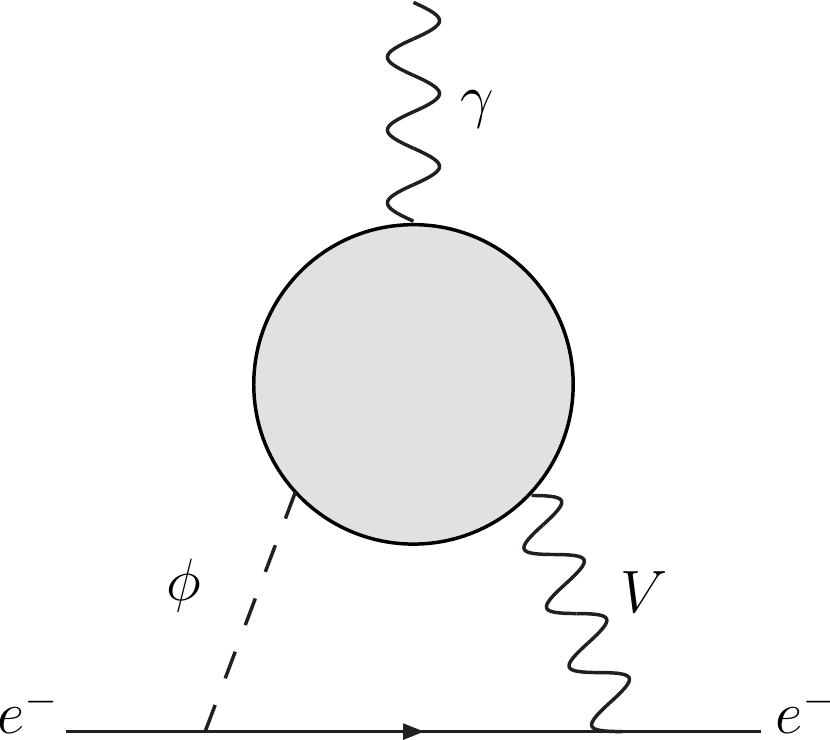}
\caption{
Barr-Zee two-loop diagram for generating eEDM,
where dominant effect is the insertion of an effective
$\phi\gamma\gamma$ coupling with top in the loop,
as in $h \to \gamma\gamma$. Here, $\phi$ can be 
any Higgs boson, with corresponding $V = \gamma,\, Z,\, W$. 
}
\label{fig:de_BZ}
\end{figure}

ACME had advanced the eEDM bound only a few years 
before~\cite{ACME:2013pal} our work, setting the impressive limit 
(compared with e.g. neutron EDM, i.e. nEDM),
\be
|d_e| < 8.7 \times 10^{-29}\; e\, {\rm cm}, \quad ({\rm ACME}\, 2014)
\label{eq:ACME14}
\ee
with the claim that the experiment knew how to improve by an order of magnitude.
Since a sizable $\rho_{tt}$ with CPV phase can affect eEDM via
the two-loop mechanism~\cite{Barr:1990vd}, i.e. the Barr-Zee diagrams
  (see Fig.~\ref{fig:de_BZ}), 
they ought to be checked.
For simplicity, we set $\rho_{ee} = 0$, i.e. we made estimates 
only with effects from $\lambda_e\,{\rm Im}\,\rho_{tt}$.
We were aware that this kind of assumption normally would require a symmetry. 
However, assuming $\rho_{ee} = 0$ in our numerics, we gave a ``predicted'' range
of  $Y_B/Y_B^{\rm obs} > 1$ that falls between Eq.~(\ref{eq:ACME14})
and the projected 
improvement, hence can be probed by ACME 
 --- with chance for discovery.

To our surprise, ACME reached their advertised improvement~\cite{ACME:2018yjb}, 
Eq.~(\ref{eq:ACME18}), in less than a year and half 
after our paper was posted~\cite{Fuyuto:2017ewj}, 
and it ruled out the entire range we had ``predicted''.

Any mechanism to render eEDM small?

{\it \textbf{Facing ThO EDM: Cancellation Mechanism}.---} \
Fortunately, the answer is yes~\cite{Fuyuto:2019svr}, 
and turned out simpler than anticipated.\footnote{
 The moral: take no shortcuts.
} 
Restoring $\rho_{ee}$,
many diagrams have to be accounted for.
The leading Barr-Zee diagram is basically inserting
$\phi \to \gamma\gamma^*$ with top loop into 
the $ee\gamma$ effective coupling,
which we call the $d_e^{\phi\gamma}$ term,
with $\phi$ summed over $h$, $H$, $A$.
Besides the $\lambda_e\,{\rm Im}\,\rho_{tt}$ term with 
respective couplings at $e$ and $t$ end, one now 
also has ${\rm Im}\, \rho_{ee}\, \lambda_t$ effects.
Interestingly, we find~\cite{Fuyuto:2019svr} a cancellation mechanism that
boils down to a simplified ``ansatz'' of
\be
\frac{{\rm Im}\, \rho_{ff}}{{\rm Im}\, \rho_{tt}} = \ r\, \frac{\lambda_f}{\lambda_t}, \ \ \
\frac{{\rm Re}\, \rho_{ff}}{{\rm Re}\, \rho_{tt}} = -\ r\, \frac{\lambda_f}{\lambda_t},
\label{eq:SM-hier}
\ee
i.e. the extra Weinberg couplings follow the mass-mixing hierarchy pattern
seen in SM couplings.
In Eq.~(\ref{eq:SM-hier}), $r$ depends on details of the loop functions involved.

\begin{figure}[t]
\center
\includegraphics[width=0.36 \textwidth]{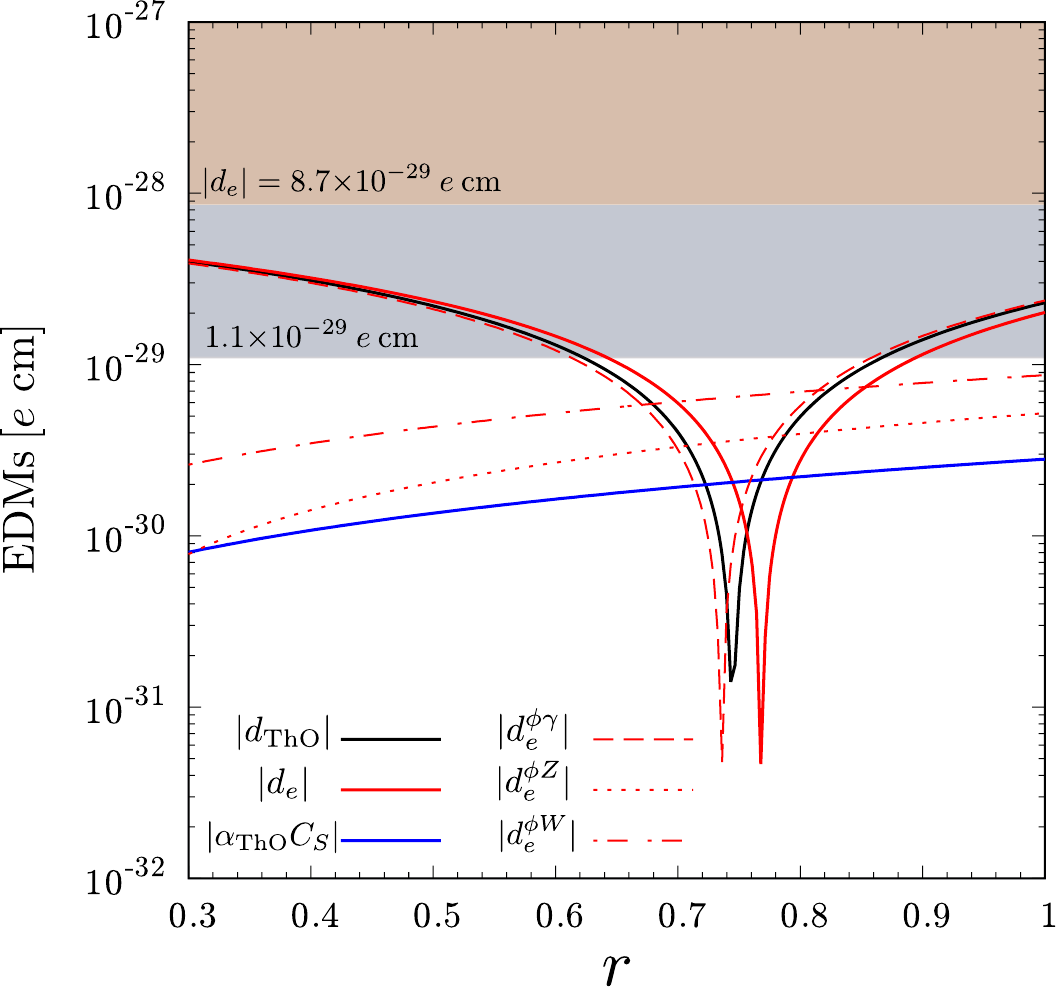}
\caption{
Electron EDM effects vs $r$, the loop functions governing two-loop Barr-Zee diagrams. 
Plot taken from Ref.~\cite{Fuyuto:2019svr}, which we refer to for details.
}
\label{fig:two-loop}
\end{figure}

We plot $|d_e^{\phi\gamma}|$ as the long-dashed line vs $r$
in Fig.~\ref{fig:two-loop}, together with the ACME\,2014 and 2018 
bounds as the two shaded regions, where the result for setting $\rho_{ee} = 0$
corresponds to $r \to 0$ and ruled out already by ACME\,2018.
Near exact cancellation occurs at $r \simeq 0.74$.
Then there are the subleading $d_e^{\phi Z}$ and $d_e^{\phi W}$ effects
 (see Ref.~\cite{Fuyuto:2019svr} for details), 
given as dot-dashed and dotted curves in Fig.~\ref{fig:two-loop},
shifting the cancellation point slightly upward
to the (red) solid curve for $d_e$. But that is not all. 
The ACME experiment actually measures the precession of eEDM 
in the strong internal electric field of the ThO molecule.
Therefore, one also has to account for~\cite{ACME:2018yjb} 
the $T$-violating $eN$ scattering effect, $\alpha_{\rm ThO} C_S$,
which involves\footnote{
 We do not give any details here.
 Suffice it to say that our results are consistent with Ref.~\cite{Cesarotti:2018huy}.
}
 all types of quarks.
Though a weaker effect than $d_e^{\phi Z}$ and $d_e^{\phi W}$,
it finally shifts $d_e$ to the (black) solid curve 
corresponding to the measured $d_{\rm ThO}$, which turns out to
lie close to the original $d_e^{\phi\gamma}$.

The result~\cite{Fuyuto:2019svr} allows a possible reduction 
by two orders of magnitude, to be probed by eEDM experiments.
Interestingly,  the cancellation mechanism of Eq.~(\ref{eq:SM-hier}) 
rhymes with the mass-mixing hierarchy that emerged 
after the Glashow-Weinberg NFC proposal.
Ref.~\cite{Fuyuto:2019svr} illustrated further the enlarged parameter space 
for EWBG allowed by this ``natural'' cancellation mechanism.

In particle physics, one is more familiar with nEDM, 
which already posed a challenge by putting strong constraints 
on BSM phases such as in SUSY.
Heavily investing on the ultracold neutron (UCN) approach, 
even just a few years back, nEDM seemed poised for a major reckoning
 (see Ref.~\cite{Chang:2017wpl} for a brief discussion).
Instead, it is the eEDM frontier that has leapfrogged forward recently,
while the nEDM front seems somewhat quiet.

ACME stands for Advanced Cold Molecule Electron EDM Experiment,
and seemingly shortened from ACMEEE.
Led by DeMille, Doyle and Gabrielse, 
this shows both cunning and self-deprecating vitality:
cunning as Wile\;E.\;Coyote, and self-deprecating in the grasp of
futility --- to capture the Road Runner, i.e. $d_e$, the grand reward.
We cannot possibly enter into technical details of 
the competitive field of eEDM experimentation,
except to say that there are many molecules, and various finesses and approaches.
For example, several years after ACME\,2014,
the JILA group~\cite{Cairncross:2017fip} more or less confirmed 
the result,
\be
d_e < 13 \times 10^{-29}\, e\, {\rm cm}, \quad ({\rm JILA}\,2017)
\label{eq:JILA2017}
\ee
using the polar molecule $^{180}$Hf\,$^{19}$F$^+$.
Amazingly, ACME jumped ahead within a year!
But the JILA result serves a useful reminder: 
the competition at $\sim 10^{-29}\, e\,{\rm cm}$ is not yet over. 
Unless a second experiment, preferably using a
different approach, confirms ACME\,2018, it is not impossible
that eEDM could still emerge around $10^{-29}\, e\,{\rm cm}$,
which {\it G}2HDM would have some preference for,
even if it could tolerate two orders of magnitude reduction.

Our result shows that  eEDM could be discovered soon,
if {\it G}2HDM is behind {\it baryogenesis}.
We have also illustrated the competition, here on {\it Earth}, 
between LEPEs and LHC in probing BSM physics, 
which is good.

So {\it G}2HDM can project: \vskip 0.1cm

\quad\quad\quad\quad\quad {\it Majestic in all the Earth;

\quad\quad\quad\quad\quad \quad Splendor above the Heavens\,!}

\subsection{\boldmath $H$, $A$, $H^+$ Spectrum: Fit for the LHC}

Let us continue with earthly pursuits at the LHC.

Although we assumed $m_H = m_A = m_{H^+} =$ 500~GeV 
for our illustration of EWBG, Fig.~\ref{fig:EWBG}, 
we have avoided facing the issue of exotic scalar mass scale so far. 
We now argue that they should be sub-TeV in mass, 
i.e. fit for the LHC.
At least the domain should be explored, rather than 
{\it assuming}~\cite{Kane:2018oax} they are much heavier 
at several TeV, as is often tacitly done now in context of SUSY.
%

Extending from the simple SM form for a single Higgs doublet $\Phi$, 
Eq.~(\ref{eq:pot}),
the Higgs potential of {\it G}2HDM, without a $Z_2$ symmetry, 
is~\cite{Davidson:2005cw,Hou:2017hiw},
\begin{align}
 &V(\Phi,\, \Phi')
= \mu_{11}^2 |\Phi|^2 +\mu_{22}^2 |\Phi'|^2
        - \left(\mu_{12}^2 \Phi^\dagger \Phi' + {\rm h.c.}\right) \notag\\
   &\ +\frac{1}{2}\eta_1|\Phi|^4 + \frac{1}{2}\eta_2|\Phi'|^4 + \eta_3|\Phi|^2|\Phi'|^2
   + \eta_4|\Phi^\dagger \Phi'|^2 \notag\\
   &\ +\left[\frac{1}{2}\eta_5^{}(\Phi^\dagger \Phi')^2
   + \left(\eta_6^{}|\Phi|^2 + \eta_7^{}
      |\Phi'|^2\right)\Phi^\dagger \Phi' + {\rm h.c.}\right]
 \label{eq:pot2}
\end{align}
where we follow the notation of Ref.~\cite{Hou:2017hiw}.
For the Higgs potential with $Z_2$ symmetry,
where $\Phi \to \Phi$ and $\Phi' \to -\Phi'$ under $Z_2$,
the usual notation~\cite{Davidson:2005cw} is $m^2_{ij}$ and $\lambda_i$, 
with $\lambda_6 = \lambda_7 = 0$.
We assume $V(\Phi,\, \Phi')$ to be $CP$ conserving for simplicity,
hence all parameters in Eq.~(\ref{eq:pot2}) are real.

{\it \textbf{\boldmath Upper Bound, 1$^{st}$EWPT, and Landau Pole}.---} \
Without $Z_2$ symmetry, $\Phi$ and $\Phi'$ cannot be distinguished
(hence the familiar $\tan\beta$ is unphysical), and 
we \emph{choose} the Higgs basis 
where $\Phi$ generates $v$, while $\mu^2_{22} >0$.
The two minimization 
conditions~\cite{Davidson:2005cw, Hou:2017hiw, Haber:2006ue, Haber:2010bw}
are
\begin{align}
  \mu_{11}^2 = -\frac{1}{2}\eta_1^{}v^2, \quad  \mu_{12}^2 =  \frac{1}{2}\eta_6^{}v^2,
 \label{eq:minim}
\end{align}
where the first is the same as in SM, 
except the factor of 1/2 due to convention.
The second minimization condition eliminates $\mu_{12}^2$ altogether, 
and $\eta_6$ is the sole parameter for $h$-$H$ mixing.
This is in contrast with usual $Z_2$ models, 
where $\lambda_6$ (and $\lambda_7$) is absent,
but requires both $m^2_{11} < 0$ and $m^2_{22} < 0$,
while the ``soft breaking'' $m^2_{12}$ parameter serves both 
roles of $h$-$H$ mixing {\it and} the decoupling mass. 
Thus,{\it G}2HDM is more intuitive than 2HDM\;I and II,
where $\mu^2_{22} > 0$ is the decoupling mass for the exotic doublet.
The exotic Higgs masses are
\begin{align}
  m_{H^+}^2 & = \mu_{22}^2 + \frac{1}{2}\eta_3^{}v^2, \label{eq:mH+} \\
  m_A^2 & =  m_{H^+}^2 + \frac{1}{2}(\eta_4^{} - \eta_{5}^{})v^2, \label{eq:mA} \\
  M_\textrm{even}^2 & =
  \left[\begin{array}{cc}
    \eta_1^{}v^2 & \eta_6^{} v^2 \\
    \eta_6^{} v^2 & m_A^2 + \eta_5v^2
    \end{array}\right],
 \label{eq:Meven}
\end{align}
%
where the last equation is for the $CP$-even Higgs sector,
which we will return to shortly.

\begin{figure}[t]
\center
\includegraphics[width=0.475\textwidth]{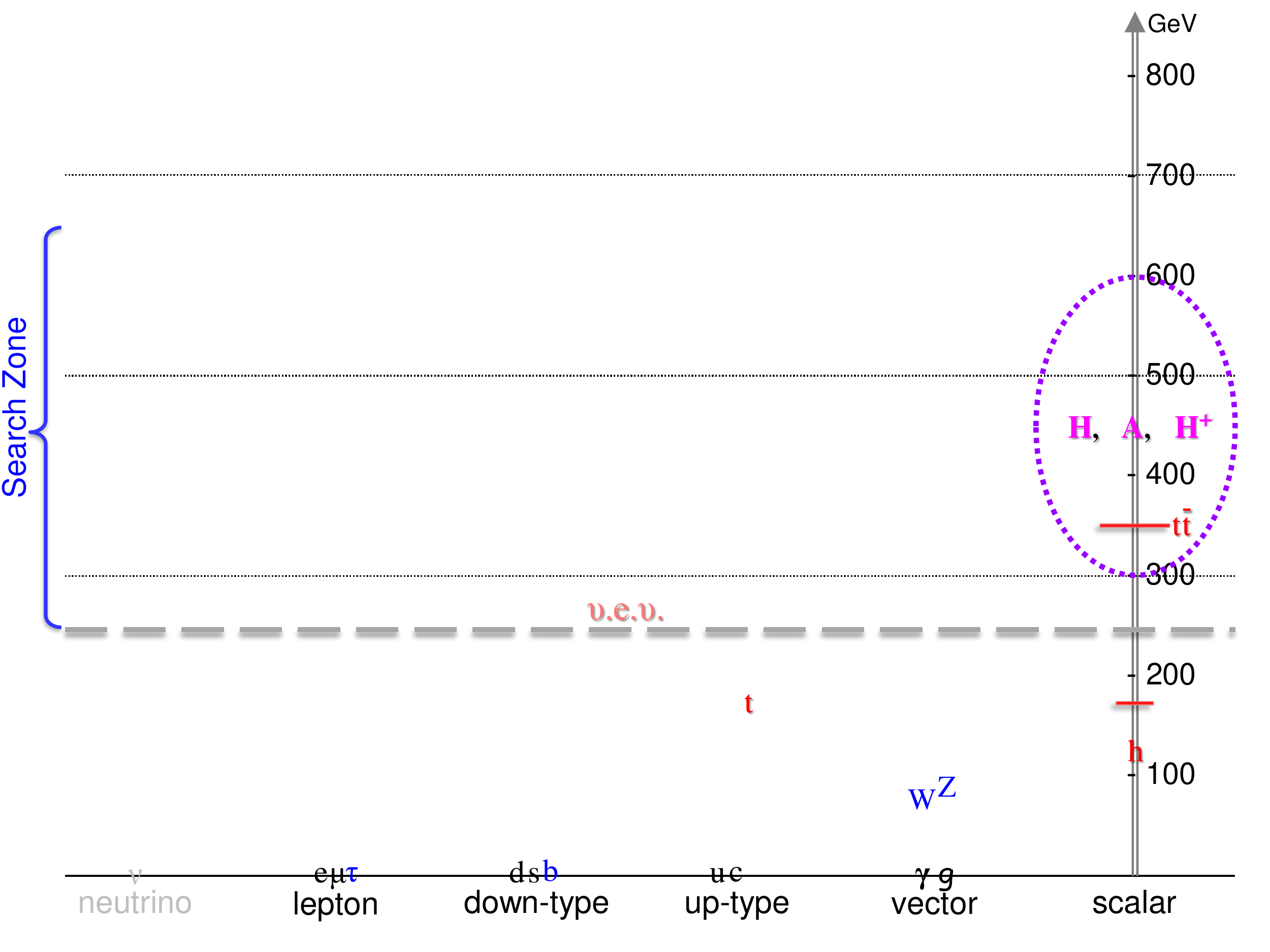}
\vskip-0.4cm
\caption{
Spectrum of matter particles, including the $t\bar t$ threshold,
with the exotic $H$, $A$, $H^+$ residing likely in the ellipse.
}
\label{fig:spectrum}
\end{figure}

With $v \simeq 246$~GeV from 
the first minimization condition of Eq.~(\ref{eq:minim}), 
the dimensionless Higgs quartic couplings, $\eta_i$ for $i = 1$ and 3--6
contribute to $m_S^2$ ($S = H, A, H^+$)  modulo $v^2$.
If we demand that $v$ sets the scale for $V(\Phi,\Phi')$, 
then all $\eta_i$s plus the dimensionless ratio $\mu^2_{22}/v^2$ 
being $O(1)$ is the common sense ``naturalness''.\footnote{
 Of course,  $\mu^2_{22}$ is an independent scale parameter.
 However, before going to the decoupling limit, one should certainly
 probe $\mu^2_{22} = {\cal O}(v^2)$. Otherwise,
 the decoupling limit of $\mu^2_{22} \gg v^2$ would 
 render all dynamical effects mute.
}
Imposing this ``naturalness'', we find
\be
m_S \lesssim 600\;{\rm GeV}. \quad (S = H, A, H^+)
\label{eq:up_bound}
\ee
i.e. for $m_S$ to go beyond 500~GeV or so 
would require some $\eta_i$, or the decoupling ratio 
$\mu^2_{22}/v^2$, to be greater than 3 or so 
(see Ref.~\cite{Hou:2017hiw} for numerical illustrations).

We remark with interest that, with $v$ as the scale parameter, 
to have all $\eta_i$s at ${\cal O}(1)$ is just what is needed 
for generating 1$^{\rm st}$EWPT~\cite{Kanemura:2004ch}, 
one of the prerequisites for EWBG. As already stated,
$\mu^2_{22}/v^2 < {\cal O}(1)$ is also needed,
otherwise decoupling would damp all scattering amplitudes,
making EWBG infeasible.
Thus, we find a {\it natural} argument for having 1$^{\rm st}$EWPT 
in {\it G}2HDM.

Having the heuristic upper bound of Eq.~(\ref{eq:up_bound}),
we will later argue a lower bound (see Fig.~\ref{fig:spectrum}) 
from the apparent smallness of $h$-$H$ mixing.
Here we note that, having $\eta_i = {\cal O}(1)$ for $i = 1$--7,
the ``common sense'' naturalness\footnote{
 For Weinberg couplings, only $\lambda_t \cong 1$ seems ``natural'',
 while the emergent mass hierarchy of Eq.~(\ref{eq:mass-hier})
 is part of the flavor riddle. 
}
has its implications, namely the {\it Landau pole} associated with scalar fields.
Even with $|\eta_i| = 1$ for some $i$ implies the 
Landau pole is not far above 10~TeV~\cite{Hou:2017hiw}.
Thus, ${\cal O}(1)$ quartics imply a strong coupling scale at several 10~TeV 
and would be a boon to a 100 TeV $pp$ collider.\footnote{
 In principle, SUSY may get a second lease of life.
}
Since none of the $\eta_i$s have yet been measured and 
some may be negative, the Landau pole scale is highly uncertain,
which we put at 10--20~GeV.

To understand the 1$^{\rm st}$EWPT and probe 
the Landau pole scale of the {\it G}2HDM Higgs potential, 
numerical studies with lattice field theory are needed, 
which could go hand in hand with experimental search.

\begin{figure}[t]
\center
\includegraphics[width=0.35 \textwidth]{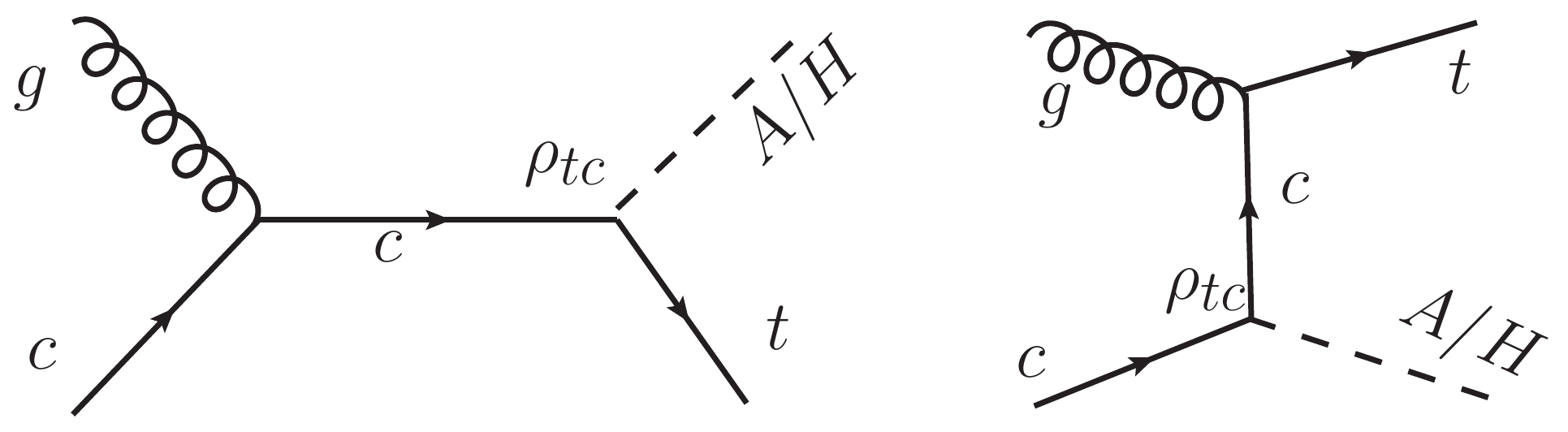} \hskip 0.5cm
\caption{
Diagram for
   $cg \to tA/H$, with $A/H \to t\bar c$, $t\bar t$ 
   leading to $tt\bar c$ and $tt\bar t$ signatures.
}
\label{fig:tAtH}
\end{figure}

{\it \textbf{Leading Search Modes at the LHC}.---} \
Shortly after the EWBG work~\cite{Fuyuto:2017ewj} 
and Ref.~\cite{Hou:2017hiw}, we proposed~\cite{Kohda:2017fkn}
the search modes via $tS$ associated production,
\begin{align}
cg \to tA/H \to tt\bar c, \quad & ({\rm Same\mbox{-}Sign\, Top})
\label{eq:SStj} \\
cg \to tA/H \to tt\bar t, \quad & ({\rm Triple\mbox{-}Top})
\label{eq:Tri-top}
\end{align}
where production goes through $\rho_{tc}$ (see Fig.~\ref{fig:tAtH}),
while $A/H \to t\bar c,\, t\bar t$ decays go through
$\rho_{tc}$ and $\rho_{tt}$, respectively,
likely the two largest extra Weinberg couplings that
can also serve as drivers~\cite{Fuyuto:2017ewj} for EWBG.
The Same-Sign Top signature, SS2$t$+$j$, where the extra jet is 
useful as a discriminant, is promising already~\cite{Kohda:2017fkn} with full Run~2 data,
while adding Run~3 data can only help.

The triple-top signature is more exquisite, 
with SM cross section only at a few fb$^{-1}$~\cite{Barger:2010uw}, 
and can allow more scrutiny.
But high statistics is needed, hence more suited~\cite{Kohda:2017fkn} 
for the High Lumi LHC (HL-LHC).
A source of frustration is the dedicated search
for $tt\bar t\bar t$ by both ATLAS and CMS, 
targeting the SM cross section~\cite{Barger:2010uw} 
at $\sim 12$\;fb$^{-1}$, which is considerably larger than tri-top in SM.
So, we had to use $4t$ search results, both Control Regions (CR)
and Signal Regions (SR), to extract~\cite{Hou:2019gpn} 
some constraint on tri-top parameter space.
Understandably, $4t$ search has a ``measurable'' SM target, 
but it amazes us that, with $tt\bar t$ cross sections that could be
over a hundred times larger than $4t$ in SM,  no dedicated search exists so far.
Philosophically, one could say that, shooting for the small but accessible
SM $4t$ process, one should be able to more or less cover 
large enhancement of tri-top, which is the basis of Ref.~\cite{Hou:2019gpn}.
The latest $4t$ search of CMS~\cite{CMS:2019rvj} and ATLAS~\cite{ATLAS:2021kqb}, 
both based on full Run~2 data, are consistent with SM, 
with ATLAS a little on the high side. Adding future Run~3 data should seal it.

It took us a while to consider $H^+$ effects,
limited in part by intuition arising from the familiar 2HDM\;II.
Motivated by Belle progress~\cite{Belle:2019iji} and Belle\;II prospects, 
in studying the $B \to \mu\bar\nu$ decay,
we finally noticed~\cite{Hou:2019uxa} the $H^+$ coupling 
in Eq.~(\ref{eq:Wein}) has its own features. 
A $b \to u$ process on the quark side, it involves two possibilities:
$\sum_i \rho_{ib} V_{ui}$, summing over $d$-type quarks,
one can show~\cite{Hou:2019uxa} it is highly suppressed; 
$\sum_i \rho_{iu}^* V_{ib}
 = \rho_{tu}^* V_{tb} +  \rho_{cu}^* V_{cb} +  \rho_{uu}^* V_{ub}
 \cong \rho_{tu}^* V_{tb}$, 
where $\rho_{cu}$ is highly constrained by $D^0$ mixing, 
and  $\rho_{uu}$ is suppressed by mass-mixing hierarchy, 
and each receive further CKM suppression. 
We noticed that, on one hand there is no $\rho_{tu}$ coupling in 2HDM\;II,
on the other hand, the associated $V_{tb}$ stands against the (2HDM\;II) 
intuition that $V_{ub}$ governs the process.
On lepton side, the $\bar\nu$ escapes Belle undetected, 
so it could be $\bar\nu_\tau$, hence bring in $\rho_{\mu\tau}$.
The upshot is that $B \to \mu\bar\nu$ decay probes the FCNH product,
\be
 \rho_{tu}\rho_{\tau\mu},
\label{eq:Bmunu-probe}
\ee
and receives an astonishing $|V_{tb}/V_{ub}|$ enhancement!

Flavor considerations matter, and one should put away old intuitions,
which are nothing but prejudice.

\begin{figure}[t]
\center
\includegraphics[width=0.28 \textwidth]{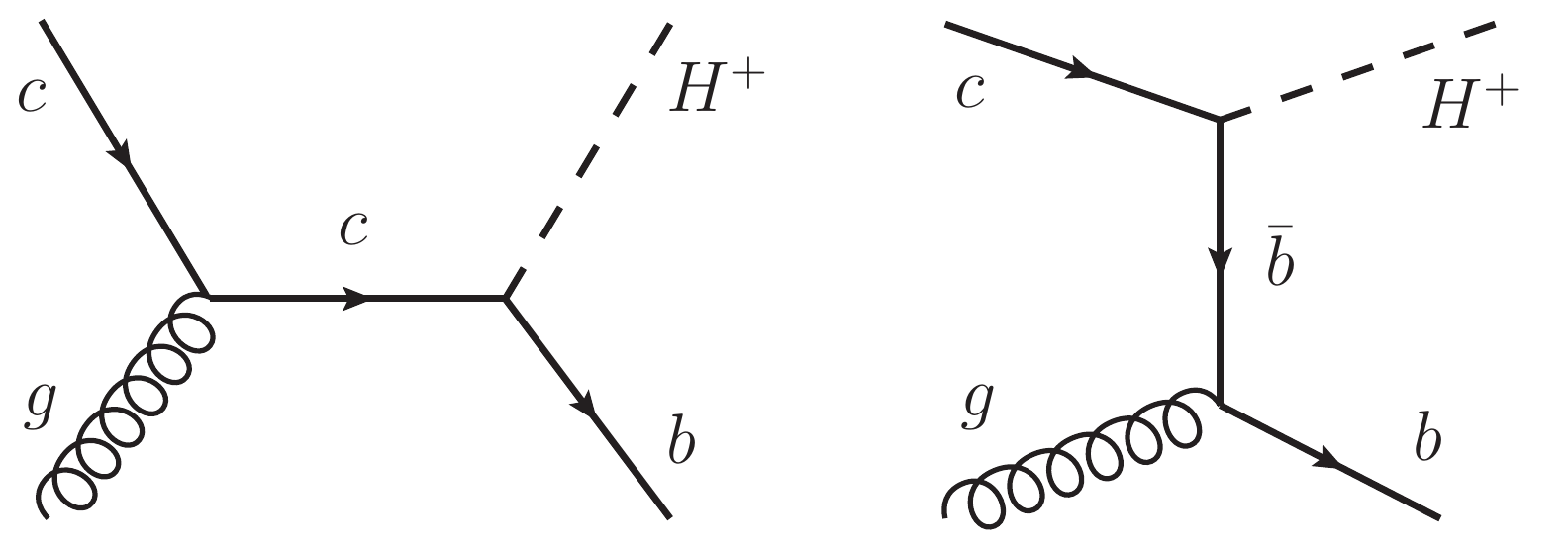}
\caption{
Diagram for
   $cg \to bH^+$, with $H^+ \to t\bar b$ 
   leading to $bt\bar b$ signature.
}
\label{fig:bH+}
\end{figure}

Gaining this insight, we proposed~\cite{Ghosh:2019exx} the search mode 
via $bH^+$ associated production (see Fig.~\ref{fig:bH+})
\begin{align}
cg \to bH^+ \to bt\bar b. \quad ({\rm Single\mbox{-}Top\,\mbox{+}}\,b\bar b)
\label{eq:btb}
\end{align}
The production is through $\rho_{tc}$,
but from the $B \to \mu\nu$ insight,
it is modulated by $V_{tb}$, hence enhanced 
by $|V_{tb}/V_{cb}|$ at amplitude level.
Because $b$ is so much lighter than $t$,
the $cg \to bH^+$ process is more efficient.
The signature of single top plus two $b$-jets may look a little dubious,
but our study~\cite{Ghosh:2019exx} did not find any obvious issue.

Our proposed production and decay processes
are a bit unconventional in the context of 2HDM, where most
intuition is from 2HDM\;II, which has SUSY as support. 
But in these $Z_2$ models, the ``attractive'' thing is
that there is practically one parameter, $\tan\beta$, i.e.
the ratio of v.e.v.s between the two doublets, aside from $H, A, H^+$ masses.
In contrast, the extra Weinberg couplings in {\it G}2HDM
open up several new parameters, hence production and decay possibilities.
One expects these would reflect the observed
mass-mixing pattern seen in the Weinberg couplings of SM,
as we learned from the Cheng-Sher ansatz, as well as 
from surviving electron EDM constraint.
That is the reason we have focused more on $\rho_{tt}$ and $\rho_{tc}$.

As a precursor to our lower mass bound on exotic scalars,
let us consider an extra pseudoscalar Higgs boson $A$.
We had worked on the $cg \to tA$ process~\cite{Hou:1997pm} 
for the LHC long before its advent.
Well into the LHC era, it is curious whether a light $A$, 
e.g. $m_A \lesssim 300$~GeV, can still be viable.
Though below $t\bar t$ threshold, one still has to 
turn off $\rho_{tt}$, as it would induce $gg \to A$, 
although one can also suppress $A \to hZ$ by small $h$-$H$ mixing
(called ``alignment'', as discussed later).
The upshot~\cite{Hou:2018zmg} is that an $A$ with mass 
between $t\bar c$ and $t\bar t$ thresholds could still survive so far,
and $cg \to tA \to tt\bar c$ of Eq.~(\ref{eq:SStj})
can probe sizable $\rho_{tc}$ at the LHC, 
accessing the second mechanism of EWBG in {\it G}2HDM.
The $\rho_{tc}$ mechanism does not generate eEDM,
hence is not constrained by ACME.

There are many other search possibilities,
depending on thresholds and other parameters.
We have worked out e.g. 
$gg \to A, H \to t\bar c$~\cite{Altunkaynak:2015twa}, $\tau\mu$~\cite{Hou:2019grj};
$cg \to tA \to tZH$~\cite{Hou:2019mve},
 where $A \to ZH$ is a weak decay;
$cg \to tH \to thh$~\cite{Hou:2019qqi} and
$cg \to bH^+ \to bhW^+$~\cite{Hou:2020tnc},
 which depend on $h$-$H$ mixing;
or $cg \to bH^+ \to bAW^+$~\cite{Hou:2021xiq}.
For these and other works, we refer to our 
brief review~\cite{Hou:2020chc} for discussion.

As far as we know, programs along some of these lines 
have started in both ATLAS and CMS.

{\it \textbf{\boldmath Remark: $A \to t\bar t$ at 400 GeV?}.---} \
With finite $\rho_{tt}$, the $H$ and $A$ bosons can be
produced via gluon-gluon fusion (ggF), $gg \to H, A$, and
it would likely decay back to $t\bar t$, 
as $\rho_{tt} = {\cal O}(1)$ is plausible.
This ``resonance'' in the $t\bar t$ channel, however,
is known to run into trouble: QCD production of $t\bar t$.
The enormous cross section of the latter means the loop-produced ggF process 
would be like foam on the waves of the wide ocean.
The interference turns the resonance peak into 
a peak-dip structure~\cite{Carena:2016npr},
making the analysis rather difficult.\footnote{
 One would think $gg \to H, A \to t\bar c$~\cite{Altunkaynak:2015twa}
 can evade this interference problem.
 However, the process may suffer from mass reconstruction.
 An excited top search~\cite{CMS:2017ixp} by CMS suggests 
 the $tj$ mass resolution at 200~GeV, which
 makes resonance search moot.
}
But ATLAS pioneered the way with 8~TeV data~\cite{ATLAS:2017snw},
finding no hint.

For some time, CMS gave no results.
But this changed in 2019, first with a somewhat ``oblique'' 
CMS PAS~\cite{CMS:2019lei}. 
``Oblique'' because no sensational words appeared 
in the Abstract, Introduction, nor Summary, although in the text, 
the word ``excess'' was used more than a few times;\footnote{
 We found out the hint of ``excess'' only when trying to incorporate the result
 into Ref.~\cite{Hou:2019gpn}. 
}
so nobody noticed, nor cared, even by summer conferences.
Words finally broke through in the paper~\cite{CMS:2019pzc}: 
in essence ``a signal-like {\it  excess} for the pseudoscalar hypothesis
  (largest) at 400\,GeV, 4\% of $\Gamma_{\rm tot}$,
  with 3.5$\sigma$ (1.9$\sigma$) local significance
  (with look-elsewhere-effect).''
However, though plots were given, no extracted values for 
coupling modifiers $g_{Ht\bar t}$ and $g_{At\bar t}$
 --- corresponding to our $|\rho_{tt}|/\lambda_t$ (real values were assumed) ---
were provided.
The result was based on $\sim 36$ fb$^{-1}$ data at 13~TeV.
CMS has not yet updated with full Run~2 data,
while ATLAS has not revealed {\it any} 13~TeV result on this so far.

How seriously should one take this hint of excess,
in a difficult process for analysis?
One could throw a host of problems at the result,
the chief one being the proximity to $t\bar t$ threshold.
The earlier ATLAS analysis~\cite{ATLAS:2017snw} 
took $m_{t\bar t} >$ 500~GeV in part to avoid such issues.
We will leave experimental issues to the experiments.

Incorporating a discussion on $A \to t\bar t$ in our ``$4t$ on $3t$'' paper,
Ref.~\cite{Hou:2019gpn} argued that having both $\rho_{tt}$ and $\rho_{tc}$
at $\simeq 1$ strength can in principle explain the excess. 
We also noted that {\it complex} $\rho_{tt}$ values 
``rotate'' in $H$-$A$ space~\cite{Hou:2018uvr},
e.g. $H$ could feign as $A$ for purely imaginary $\rho_{tt}$
(hence the analysis should be done with complex coupling modifiers).
The need for $\rho_{tt}, \rho_{tc} \simeq 1$ is in part to reduce 
the decay branching ratios, but such large strength for the relatively
low $m_A$ causes other tensions. The proposed 
$cg \to bH^+$ production~\cite{Ghosh:2019exx} brings in new worries, 
e.g. sizable cross section for relatively high $H^+$ masses.
If $H^+ \to AW^+$ is allowed~\cite{Hou:2021xiq},
followed by $A \to t\bar c$, $t\bar t$, one would face $4t$ bounds.
So, we find the hint for a 400~GeV psuedoscalar by CMS dubious.
However, such $t\bar t$ resonance searches should be watched,
while the experiments should provide the coupling modifier values.

\subsection{Nature's Flavor Design and Lower Mass Bound}

When we used the Gell-Mann Principle to demand the existence
of a second Higgs boson doublet $\Phi'$, implying the existence
of exotic Higgs bosons $H$, $A$ and $H^+$, it is not particularly new, 
but we raised two related questions:

\quad ``Where are they?''

\quad ``What do they do?''

\noindent These lead to a combined question:
\be 
\mbox{\it ``What\, hides\, $H, A, H^+$\, effects\, from\, our\, view?''}
\label{eq:hideS}
\ee

For 2HDM\;II, there are specific predictions for exotic scalar couplings,
governed by $\tan\beta = v_1/v_2$, the ratio of v.e.v.s of the two doublets.
One could enhance $d$-type quark and charged lepton $\ell$ couplings,
or enhance $u$-type quark couplings, including the dominant top,
or treat them on more equal footing, i.e. $\tan\beta \sim 1$.
For a long time, $H^+$ search was for $m_{H^+} < m_t$,
and only more recently do we have search bounds for 
$pp \to \bar tH^+(b)$ from CMS~\cite{CMS:2020imj} 
and ATLAS~\cite{ATLAS:2021upq}. 
The parton level process is
\be
\bar bg \to \bar tH^+,
\label{eq:tH-}
\ee
which can be compared with our $cg \to bH^+$ process,
Eq.~(\ref{eq:btb}) or Fig.~\ref{fig:bH+}.
In {\it G}2HDM, Eq.~(\ref{eq:tH-}) is governed by $\rho_{tt}V_{tb}$,
while Eq.~(\ref{eq:btb}) is governed by $\rho_{tc}V_{tb}$,
and are quite independent, the latter receiving $|V_{tb}/V_{cb}|^2$
enhancement~\cite{Ghosh:2019exx} in rate compared with 2HDM~II, 
while phase space favored.
It also implies that $H^+ \to c\bar b$ proceeds with 
$\rho_{tc}V_{tb}$ coupling and can compete strongly 
with $H^+ \to t\bar b$, which is quite unconventional. 
The measurement of $\bar tH^+(b)$ production,
taking the $H^+ \to c\bar b$ dilution into account, 
has been used to constrain the extra Weinberg couplings.
However, CMS has yet to update with full Run~2 data.
Both experiments have~\cite{CMS:2020imj, ATLAS:2021upq} 
a more stringent than expected bound at 600--700~GeV, 
putting some pressure on heavy $H^+$.

{\it \textbf{Nature's Flavor Design}.---} \
So, we have illustrated that, similar to
$H, A \to t\bar c$ and $t\bar t$ can mutually dilute each other,
$H^+ \to c\bar b$ and $ t\bar b$ decays can also mutually dilute each other, 
and should be taken into account. 
This is part of the broader ``flavor design'' of Nature
that addresses the question in Eq.~(\ref{eq:hideS}).

Let us articulate once more:
\begin{enumerate}
 \item (Fermion) Mass-Mixing Hierarchy

   The fermion mass hierarchies,
     Eqs.~(\ref{eq:mass-hier}) and (\ref{eq:toverb}),
   and mixing hierarchy Eq.~(\ref{eq:mixing-hier}), all {\it emergent},
   seem to have worked in concert such that we have been oblivious
   to extra Weinberg couplings,
\be
\mbox{\it $\rho_{ij}^f$\;trickle\;off\;going\;off-diagonal,}
\label{eq:trickle}
\ee
   as originally noted by Cheng and Sher~\cite{Cheng:1987rs}.
   These hierarchies form the flavor enigma,
   which we do not understand, but seem to be Nature's choice,
   as emphasized by the word {\it emergent}.
 \item ({\it Emergent}) Alignment

  Another phenomenon, not quite related to fermion flavor,
 {\it emerged} after the discovery of $h(125)$: 
\be
  \quad\quad\quad\ \  \mbox{\it The\;$h$\;boson\;resembles\;$H_{\rm SM}$\;rather\;closely.}
\label{eq:alignment}
\ee
  So far we
  have not seen deviations in its properties from $H_{\rm SM}$,\footnote{
    This emerged shortly after discovery of $h(125)$, and became
    firm with the ATLAS-CMS combination of Run~1 data~\cite{ATLAS:2016neq}.
    ATLAS recently updated with full Run~2 data~\cite{ATLAS:2020qdt},
    affirming the case.
}
  the Higgs boson of SM. If there is a second Higgs doublet
    --- {\it by Gell-Mann Principle there must be!}
  --- then $h$--$H$ mixing is small,
  i.e. the mixing angle $c_\gamma \equiv \cos\gamma$
  (called $\cos(\beta - \alpha)$ in 2HDM~II convention)
  between the two $CP$-even scalars is small. 
  This {\it alignment} phenomenon can readily account for
  the absence of $t \to ch$ and $h \to \tau\mu$ so far.
  That is, Eqs.~(\ref{eq:tch}) and (\ref{eq:htamu}) can be
  satisfied without requiring e.g. $\rho_{tc}$ to be small.

  Historically, whether in the Cheng-Sher ansatz~\cite{Cheng:1987rs}
  or our proposal~\cite{Hou:1991un} of $t\to ch$
  (alternatively, $h \to t\bar c$) in ``2HDM~III'',
  Higgs sector mixing was glossed over, reflecting the expectation
  that the 5 Higgs bosons would share some common spectrum,
  and would be well-mixed.
  As it now stands, and as we have illustrated in Sec.~IV.B,
  there seems some gap between $h$ at 125~GeV
  and the exotic $H, A, H^+$ bosons, and Nature threw in further
  a small $c_\gamma$ to protect $h$ from FCNH couplings.

  \quad\ \underline{NFC of Glashow-Weinberg should be retired.}
  
  We elucidate alignment further below to extract some
  lower bound on exotic Higgs boson masses, 
  but turn now to state a curiosity.
\item Near diagonal {\boldmath $\rho$}$^d$: \
  Nature's mysterious ways 

  In the SM context, 
  $K^0$--$\overline K^0$, $B^0$--$\overline B^0$ and
  $B_s^0$--$\overline B_s^0$ systems are our most sensitive probes of
  flavor and $CP$ violation. But absence of New Physics so far in these sectors
  (bounds are still suitably accommodating), as well as in e.g. $B_q \to \mu^+\mu^-$,
  means that the {\boldmath $\rho$}$^d$ matrix is near-diagonal.

  Numerically, it is not that serious, e.g. $\rho_{bs}$ only needs to be
  an order of magnitude smaller than $\rho_{bb}$, which is already constrained
  by~\cite{Altunkaynak:2015twa} $b \to s\gamma$ --- another sensitive probe ---
  to be small, $\rho_{bb} \lesssim \lambda_b \sim 0.02$.
  However, were it not the case, that Nature put in 
  just a tad larger off-diagonal elements in $\rho_{ij}^d$,
  we would have detected the effects via these exquisite systems,
  and discovered BSM much earlier.
  Instead, all along we just confirmed SM, from
  the fantastic working of GIM,
  to the present day lack of firm BSM effects.
\end{enumerate}

{\it \textbf{Alignment \& Lower Mass Bound}.---} \
Let us turn to some subtlety of {\it alignment} in {\it G}2HDM.
From Eqs.~(\ref{eq:mH+})--(\ref{eq:Meven}),
we had inferred an upper bound of Eq.~(\ref{eq:up_bound})
by argument of ${\cal O}(1)$ Higgs quartics and $\mu^2_{22}/v^2$,
the decoupling parameter,
also needed for achieving 1$^{\rm st}$EWPT.
By the same token, we argue that one has a rough lower bound on $m_H$,
or the generic $m_A$.

Diagonalizing Eq.~(\ref{eq:Meven}) by
\begin{align}
  R^T_\gamma M_\textrm{even}^2 R_\gamma  =
    \left[\begin{array}{cc}
    m_H^2 & 0 \\
    0 & m_h^2 \\
  \end{array}\right], \,\
  R_\gamma  =  \left[\begin{array}{cc}
    c_\gamma & - s_\gamma \\
    s_\gamma & c_\gamma \\
  \end{array}\right],
\end{align}
where $s_\gamma \equiv \sin\gamma$.
In 2HDM~II convention, $\gamma = \beta - \alpha$
is the relative angle between the Higgs basis and
the neutral Higgs mass basis, hence basis-independent.
Fixing $m_h \cong$ 125~GeV, 
without solving explicitly for $m_H$, 
the mixing angle $c_\gamma$ satisfies two relations~\cite{Hou:2017hiw},
\begin{align}
   c_\gamma^2  = \frac{\eta_1^{}{v^2} - m_h^2}{m_H^2 - m_h^2}, \quad 
  \sin2\gamma^{}  = \frac{2\eta_6^{}v^2}{m_H^2 - m_h^2}.
 \label{eq:c_ga-s_ga}
\end{align}
Since we know that $c_\gamma$ is small, 
i.e. approximate alignment~\cite{ATLAS:2016neq},
$s_\gamma \to -1$ much faster than $c_\gamma \to 0$, 
hence the second relation simplifies to
\begin{align}
  c_\gamma^{} & \simeq \frac{-\eta_6^{}v^2}{m_H^2 - m_h^2},
  \ \ \ {\rm (approx.\; alignment)}
 \label{eq:Haber}
\end{align}
which is the familiar mixing formula for a two-level system.
Since $m_h^2$ is small compared with $v^2$
 (roughly $v^2/4$ in the present convention),
we see that for $m_H^2 \lesssim v^2$,
small $c_\gamma$ can only be achieved by having $\eta_6$ small.

In fact, Ref.~\cite{Hou:2017hiw} was motivated by
the statement that ${\cal O}(1)$ Higgs quartics are 
needed for 1$^{\rm st}$EWPT~\cite{Kanemura:2004ch}.
But from the SUSY mindset,
$\eta_6$ appeared to be small~\cite{Bernon:2015qea, Bechtle:2016kui}. 
Concerned with logical consistency
 (one quartic kept small defeats reasoning of generic ${\cal O}(1)$ quartics), 
the thought was to investigate how large $\eta_6$ could be, 
and we found that $\eta_6$ at ${\cal O}(1)$ was truly allowed.

\begin{figure}[t]
\center
\includegraphics[width=5.5cm]{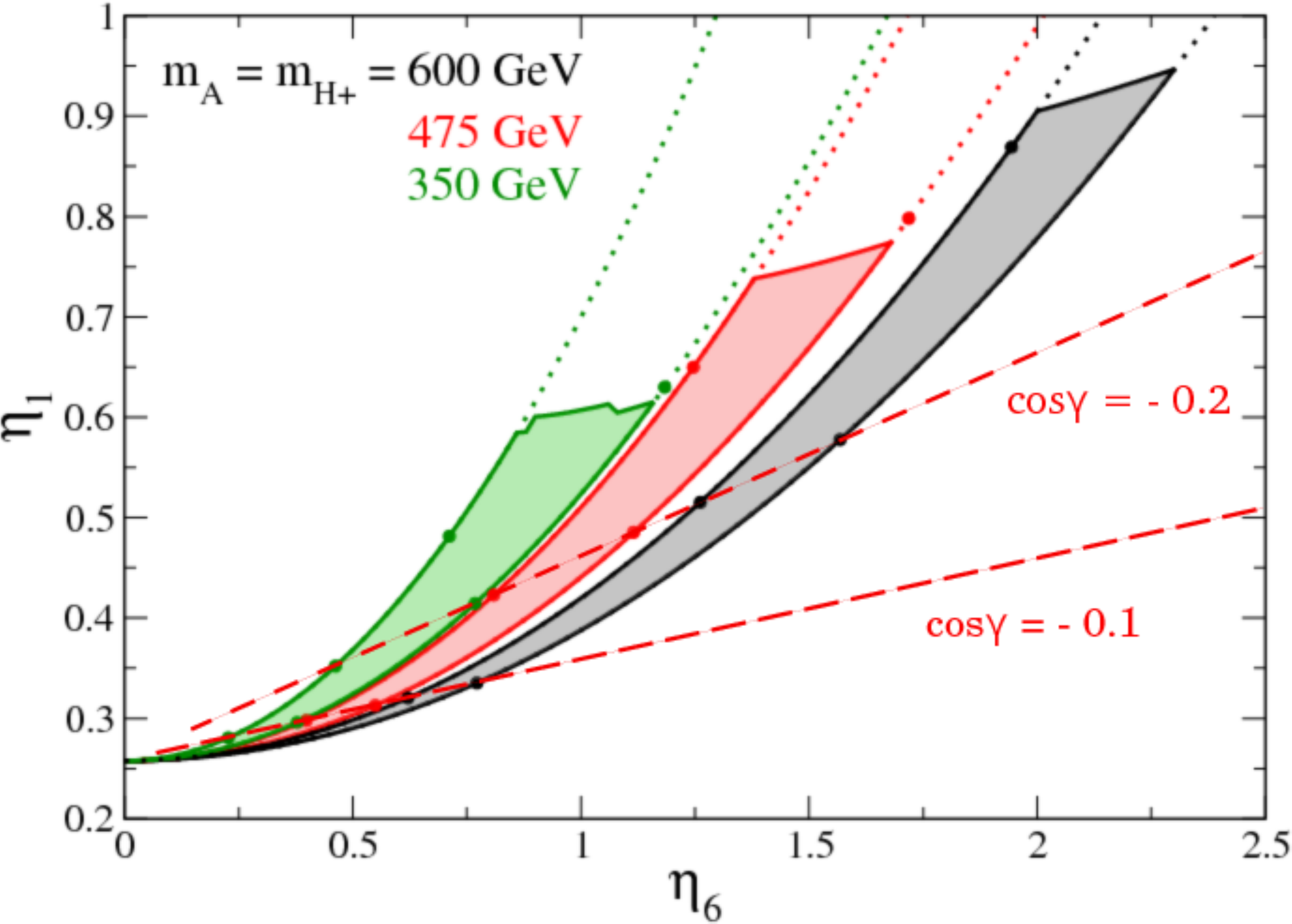}
\caption{
 Higgs quartic couplings $\eta_1$ vs $\eta_6$, where
 bullets are for $-\cos\gamma = 0.1,\ 0.2,\ 0.3$
 ($-\sin\gamma \cong 0.995,\ 0.980,\ 0.954$).
 [Figure taken from Ref.~\cite{Hou:2017hiw}.]
}
\label{fig:eta6_eta1}
\end{figure}

Let us illustrate~\cite{Hou:2017hiw} some simple but subtle aspects
of alignment in {\it G}2HDM, free from any SUSY distraction.
For $m_A = m_{H^+} = 350, 475, 600$\;GeV, 
i.e. custodial symmetry to control $T$-parameter 
and implemented by $\eta_4 = \eta_5 \in (0.5, 2)$, 
hence $m_H$ is heavier than $m_A = m_{H^+}$,
we plot in Fig.~\ref{fig:eta6_eta1}
the two controlling quartics for $h$--$H$ mixing, $\eta_1$ vs $\eta_6$. 
In Fig.~\ref{fig:eta6_eta1}, bullets are 
for $-\cos\gamma = 0.1,\ 0.2,\ 0.3$, where two sets are 
connected by (red) dashed lines.
The unmarked line for $c_\gamma = -0.3$ roughly coincides 
with the 95\% C.L. limit from $S$-$T$ data~\cite{Baak:2012kk},
which cuts off the parameter space of each ``horn'', turning into dotted lines;
small $c_\gamma$ and $\Delta T$ constraint seem correlated.

To illustrate ${\cal O}(1)$ quartics, let us arbitrarily
take the middle (pink) ``horn'', and $(\eta_6,\, \eta_1) \simeq (1.25,\, 0.6)$.
Both values are clearly ${\cal O}(1)$, with $c_\gamma \sim -0.25$.
One can approach $|c_\gamma| \simeq 0.2$ by picking 
a larger value of $\eta_6 \simeq 1.5$ in the $m_A = m_{H^+} = 600$~GeV
(grey) ``horn''.
So how is this achieved?
Note that the $\eta_1$ value of $\sim 0.6$ is 2.3 times the value
of $\sim 0.26$ that corresponds to SM value.
This {\it enhanced $hhh$ coupling} itself would have
implications for the electroweak phase transition.
But now we see that the eigenvalue of $m_h \simeq$ 125~GeV 
is actually maintained by the familiar {\it level repulsion}:
the lower state is pushed down to 125~GeV,
while the large $\eta_6$ mixing quartic pushes the second state up
to a higher $m_H$ value, and helps suppress $c_\gamma$ through 
the denominator of Eq.~(\ref{eq:Haber}).

The upshot is that there is large parameter space,
the body of the ``horns'', rather than the ``tip of the horn''
where all ``horns'' converge, 
as $\eta_6$ necessarily shrinks with $c_\gamma$,
with $\eta_1 v^2 \to m_h^2$ as required by 
the first part of Eq.~(\ref{eq:c_ga-s_ga}),
in the approach to the {\it alignment limit}.

If having Higgs quartics ${\cal O}(1)$ demand $\eta_6 > 0.3$,
then the bulk of the horn roughly demands
\be
 m_H \gtrsim v \approx 250\,{\rm GeV}.
\label{eq:low_bound}
\ee
There is a catch, though.
Eq.~(\ref{eq:low_bound}) is not fool proof,
as Nature can impose a small $\eta_6$ 
when the rough bound of Eq.~(\ref{eq:low_bound}) is not kept, at her will.
What may be unclear is whether 1$^{\rm st}$EWPT can be maintained.
But that can arise from other means, at Nature's will.

\subsection{Glimpse of the New Flavor Era}

We have three new $3 \times 3$ extra Weinberg matrices,
each element in principle complex, so altogether 54 new flavor parameters.
It may seem horrible to have {\it so many} new parameters
in {\it G}2HDM (but don't forget the many flavor parameters of SM). 
On the other hand, it is all the more remarkable that
we have not seen hints of their existence so far
 --- the {\it Flavor Design} of Sec.~IV.C!

Actually, there have been a host of suitably strong hints from 
flavor sector for BSM physics, the so-called ``$B$-anomalies''. 
In order of their first appearance (none are viewed as established by itself):
 $R_D, R_{D^*}$ anomaly;
 $P^\prime_5$ anomaly;
 $R_K, R_{K^*}$ anomaly.
Although people tried {\it G}2HDM with $R_D, R_{D^*}$ anomaly,
but it was insufficient.

\begin{figure*}[t]
\center
\includegraphics[width=10.5cm]{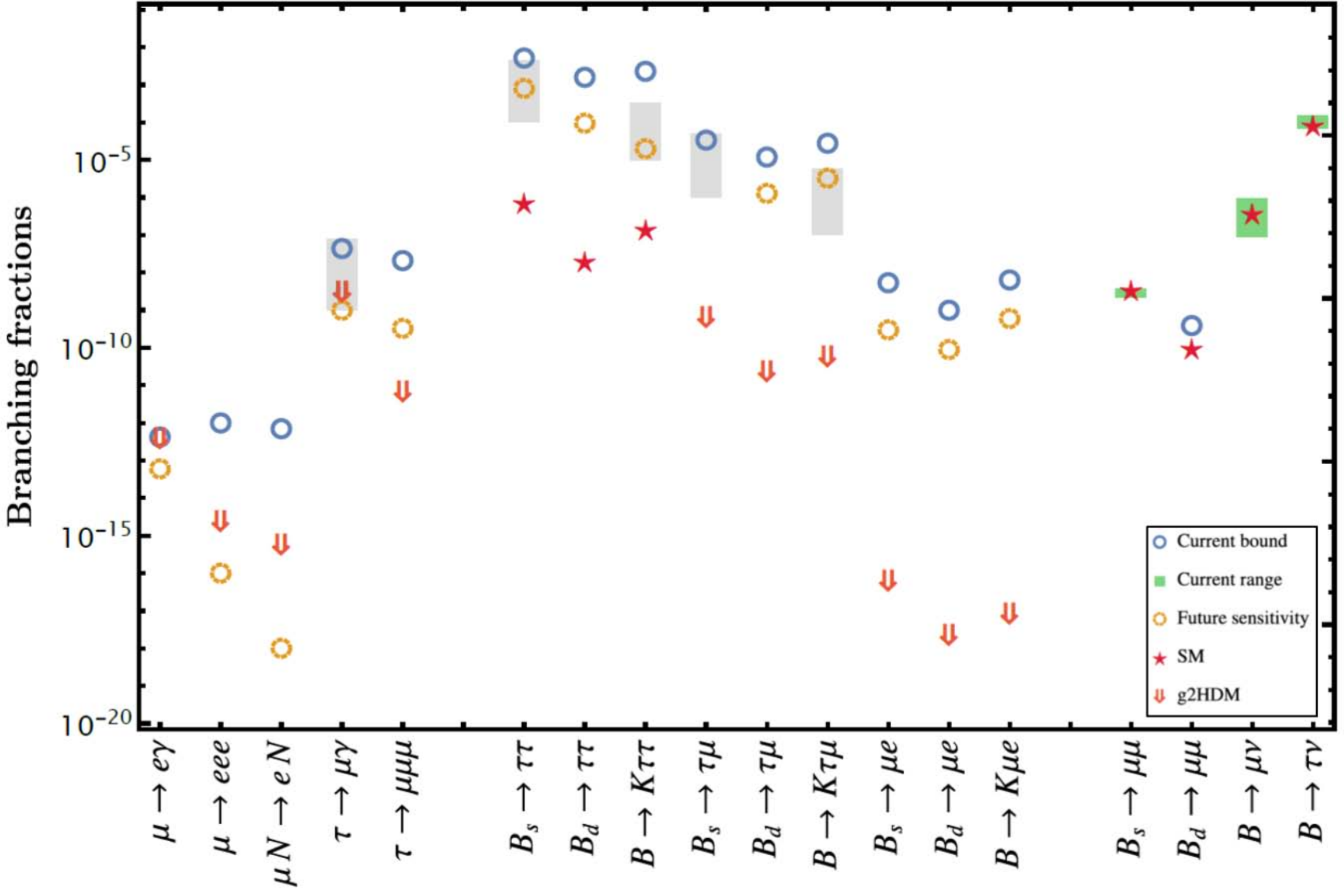}
\caption{
  Important flavor measurements to watch, with
  blue solid circles for current bounds,
  orange dotted circles for future sensitivities,
  green shaded bands for the measured ranges of
   $B_s \to \mu\mu$ and $B \to \tau\nu,\, \mu\nu$,
  and red $\star$ marking SM predictions,
  while red $\Downarrow$ illustrate {\it G}2HDM benchmark projections,
  where we use $c_\gamma = 0.05$, $m_{H, A} = 300$~GeV,
  $\rho_{\mu e} = \lambda_e$, $\rho_{\tau\mu}=\lambda_\tau$,
  and $\rho_{ii} = \lambda_i$, except $\rho_{tt} = 0.4$.
  See text for further details, including the five grey shaded bands.
  [Figure taken from Ref.~\cite{Hou:2020itz}.]
}
\label{fig:LFV_sum}
\end{figure*}

%
The first $R_{D^{(*)}}$ anomaly started with BaBar~\cite{BaBar:2012obs},
which found deviations of $B \to D^{(*)}\tau\nu$ from $B \to D^{(*)}\mu\nu$.
It became a sensation when it was confirmed by 
LHCb~\cite{LHCb:2015gmp} in 2015.
The $P^\prime_5$ anomaly is from LHCb,
in the named variable in some $q^2$ bins in the angular analysis of 
$B^{0} \to K^{*0} \mu^{+} \mu^{-}$~\cite{LHCb:2015svh}.
The $R_K$~\cite{LHCb:2014vgu}, 
then $R_{K^*}$~\cite{LHCb:2017avl} anomalies are also from LHCb,
in the ratio of $B \to K^{(*)}\mu\mu$ with $B \to K^{(*)}ee$
in some choice $q^2$ range.
These ``$B$-anomalies'' are quite well known, so we do not 
describe them further, but refer to our time-capsule snapshot
of 2018~\cite{Hou:2019dgh}, where a brief discussion of each ``anomaly'', 
together with a critique from experimental perspective, is given.
Nothing has drastically changed since then.

Two BSM possibilities emerge from the $B$-anomalies, 
one is some $Z'$, another is some leptoquark (LQ), 
each accounting for a subset of the $B$-anomalies. 
We will use the LQ from ``PS$^3$''~\cite{Bordone:2017bld}, i.e.
having three copies of the Pati-Salam symmetry~\cite{Pati:1974yy}, 
as a contrast example.

One should not forget the recent result announced by
the FNAL Muon g-2 experiment~\cite{Muong-2:2021ojo}, 
confirming the previous~\cite{Muong-2:2006rrc} BNL result.
Standing alone at 4.2$\sigma$, this is the single most significant 
discrepancy. 
We will  illustrate what may happen in {\it G}2HDM
 towards the end.

We plot in Fig.~\ref{fig:LFV_sum} important flavor measurements 
to watch~\cite{Hou:2020itz} in the coming future.
We start from $\mu$FV and $\tau$FV,
then present various rare $B$ decays into final states
involving $\mu$ and $\tau$.
We take $c_\gamma = 0.05$, $m_{H,\, A} = 300$~GeV,
  $\rho_{\mu e} = \lambda_e$, $\rho_{\tau\mu}=\lambda_\tau$,
  and $\rho_{ii} = \lambda_i$, except $\rho_{tt} = 0.4$ for illustration.
That is, we follow the pattern
\begin{align}
\rho_{3j}^f  \lesssim \lambda_3^f \ \, (j \neq 1), \quad\quad\ \
\rho_{i1}^f \lesssim \lambda_1^f,
\label{eq:rho3jrhoi1}
\end{align}
that seem to conservatively reflect Nature's flavor design, 
and take moderate value for $\rho_{tt}$.

{\it \textbf{\boldmath $\mu$FV \& $\tau$FV}.---} \
As noted, particle physics started in some sense 
when nature of the muon was clarified by the $\pi$-$\mu$-$e$ decay chain.
Starting with the pioneering work of Pontecorvo~\cite{Hincks:1948vr},
MEG\,II is the current torch-bearer in the search for $\mu \to e\gamma$,
which is the first entry in Fig.~\ref{fig:LFV_sum}, and we see
that it probes new territory.
In Fig.~\ref{fig:LFV_sum}, the solid  (blue) circles
are the current bounds, usually taken from PDG~\cite{PDG},
and in this case the MEG experiment. 
MEG\,II would improve by about an order of magnitude, the (orange) dotted circle.
For {\it G}2HDM, the process arises from a two-loop mechanism
similar to Fig.~\ref{fig:de_BZ}, with incoming $e$ replaced by $\mu$,
hence the Weinberg coupling $\rho_{\mu e}$.
Taking $\rho_{\mu e} = \lambda_e$ in the plot (recall eEDM),
the two-loop mechanism is driven by $\rho_{tt}$,
where the value saturates MEG bound, illustrated by 
the (red) $\Downarrow$ benchmark projection in {\it G}2HDM.
To go beyond MEG~II sensitivity, new technology/design needs to be brought in.

A second experiment, Mu3e, searches for $\mu \to e\bar ee$.
With $\rho_{ee} = \lambda_e$, however, the tree-level effect
is too suppressed, and $\mu \to e\bar ee$ is dominated by the
$\mu e\gamma$ dipole, marked by the (red) $\Downarrow$.
Though within reach of Mu3e, measurement would come late.

A third type of experiment, $\mu \to e$ conversion on Nuclei, 
or $\mu N \to eN$, may become the champion. 
This is because of ambitious plans to improve 
on existing bound~\cite{PDG} by six orders or more,
starting with COMET at KEK and Mu2e at FNAL.
Our  (red) $\Downarrow$ in Fig.~\ref{fig:LFV_sum} may not
look too impressive. But unlike $\mu \to 3e$, 
at the nuclei side, the $(\bar e\mu)(\bar qq)$ four-Fermi
operators probe {\it all} $\rho_{qq}$s,
and the ability to use different nuclei increases the versatility.
What is important, however, is to realize the 
many orders of magnitude improvement. Thing to watch!

Turning to $\tau$ decays, the analogous modes are
$\tau \to \mu\gamma$ and $\tau \to \mu\bar\mu\mu$.
Abiding by Eq.~(\ref{eq:rho3jrhoi1}) and with 
$\rho_{tt}$ fixed by $\mu \to e\gamma$, it would take~\cite{Belle-II:2018jsg} 
close to full Belle~II data to discover $\tau \to \mu\gamma$,
while the expected dipole dominance indicates that
$\tau \to \mu\bar\mu\mu$ lies out of reach.

So, there are some prospects for discoveries in $\mu$FV and $\tau$FV processes.
But they are not guaranteed, as it depends on
$\rho_{tt}$ and $\rho_{\mu e}$ values together saturate
the current MEG bound~\cite{PDG}.
That is the reason we have the (red) $\Downarrow$,
that the {\it G}2HDM ``prediction'' lies below.

{\it \textbf{\boldmath Various Rare $B$ Decays and LQ Contrast}.---} \
With $\tau \to \mu\gamma$, we already noticed the grey shaded band
that more or less fills the space from current bound\footnote{
 The current bound~\cite{PDG} from BaBar has just been
 improved incrementally by Belle~\cite{Belle:2021ysv} with full dataset,
 which means Belle~II would continue to push the bound
 as data accumulates.
}
to the projection for Belle~II.
This is from a phenomenological study~\cite{Bordone:2018nbg} of the PS$^3$ model
due to the $b\bar\tau$-flavored leptoquark, which carry also lower generation flavors,
with the experimental result taken as constraint.

Having three copies of PS symmetry to avoid flavor constraints
would violate lepton universality, leading to LFV such as 
$B \to K\tau\mu$ and $B_s \to \tau\mu$ by general arguments~\cite{Glashow:2014iga}.
In the PS$^3$ study of Ref.~\cite{Bordone:2018nbg},
$B_s \to \tau\mu$ as large as $\sim 5 \times 10^{-4}$ was predicted.
Within a year, however, LHCb pushed the bound down~\cite{LHCb:2019ujz} 
by more than an order of magnitude, ruling out 
the bulk of the predicted region~\cite{Bordone:2018nbg}, and 
the PS$^3$ practitioners had to tweak right-handed couplings,
giving this new ``grey band''~\cite{Cornella:2019hct} 
in Fig.~\ref{fig:LFV_sum}, with current experimental 
bound~\cite{LHCb:2019ujz} biting right in.
This interaction between theory and experiment 
in $B_s \to \tau\mu$ entry\footnote{
 A fresh result on this mode just came out from Belle~\cite{Belle:2021rod},
 showing that Belle~II can compete with LHCb in this mode.
} 
is as it should be and exciting.
The reach of LHCb with given data remains to be seen,
but likely can improve with data.

From Ref.~\cite{Bordone:2018nbg} to Ref.~\cite{Cornella:2019hct},
the experimental constraint on $\tau\to \mu\gamma$ has barely moved,
awaiting Belle~II data.
For $B^+ \to K^+\mu^-\tau^+$~\cite{Glashow:2014iga}, 
the current bound~\cite{PDG} is held by BaBar~\cite{BaBar:2012azg},
which used full hadronic tag of the other $B$,
and look for events in the $m_\tau$ bin on signal side,
given that $K^+$ and $\mu^-$ are well-identified tracks.
We note with interest that LHCb has found a method for 
analogous kinematic control,
and using full data to date, reached a bound~\cite{LHCb:2020khb} 
on $B^+ \to K^+\mu^-\tau^+$ only slightly worse than BaBar.
Thus, LHCb would soon take over in pushing down the bound
to access the grey band from PS$^3$.
So this could be the smoking gun, or the way to further rule out PS$^3$.
But Belle has not yet followed up even with BaBar~\cite{BaBar:2012azg}
 in this mode!
We assume this would happen soon, and that
Belle\;II would come head to head with LHCb in the coming years.
The two other ``grey band'' modes to probe LQ effects in PS$^3$,
$B_s \to \tau\tau$ and $B \to K\tau\tau$ is less crisp than
$B_s \to \tau\mu$ and $B \to K\mu\tau$.

The discussion above is rather exciting for the discovery potential,
and we would be more than happy if it happens.
But otherwise we are no fans of LQs ({\it Why LQ Now\,!?}~\cite{Hou:2019dgh})
Turning around to look at projections in {\it G}2HDM,
we note that $B_q \to \tau\tau$
and $B \to K\tau\tau$ have SM expectations, the (red) $\star$'s
in Fig.~\ref{fig:LFV_sum}.
The interference effect with {\it G}2HDM would be similar to
$B_q \to \mu\mu$ discussed below.
For $B_q \to \tau\mu$ and $B \to K\mu\tau$,
the projection is 4--5 orders of magnitude below experimental sensitivity,
while final states with $\mu e$ are even farther away.
This is because we adhere to Eq.~(\ref{eq:rho3jrhoi1}),
and illustrates our answer to the question in Eq.~(\ref{eq:hideS}):
the mass-mixing hierarchy, alignment, and near diagonal $\rho_{ij}^d$
continue to hide {\it G}2HDM from our view in LFV $B$ decays.

{\it \textbf{\boldmath {\it G}2HDM--SM Interference Modes}.---} 
Finally, we come to several very interesting modes that can enjoy
interference with SM process, and are elevated probes of {\it G}2HDM,
namely $B_s \to \mu\mu$, $B_d \to \mu\mu$ and $B^- \to \mu\nu, \tau\nu$.

The decays $B_s \to \mu\mu$ and $B_d \to \mu\mu$ have been of 
great interest since CDF times, largely due to
a very well known $(\tan\beta)^6$ enhancement in SUSY-like settings.
Alas, as Tevatron passed the torch on to the LHC, this potentially large 
enhancement was eventually excluded by experiment.
We now have good indication~\cite{PDG} that $B_s \to \mu\mu$ is
even slightly smaller than SM expectation. while $B_d \to \mu\mu$
measurement is not yet settled. 
The front runners for measurement are LHCb and CMS.

We have commented that the smallness of $B_q \to \mu\mu$
(and lack of BSM indications in $K^0$, $B_d$ and $B_s$ meson mixings)
demands a near diagonal $\rho_{ij}^d$.
But can the SM dominance of these modes,
by interference, help us probe {\it G}2HDM?
It could happen. 
But there can be both tree or loop~\cite{Hou:2020itz} effects 
in {\it G}2HDM, and possibilities abound (too many parameters). 
How to approach it?

The {\it G}2HDM effects at tree level in $B \to \mu\nu$ and $\tau\nu$ have
also been touched upon, that subtleties involving charged Higgs effects
and the undetected neutrino go beyond the intuition from 2HDM\;II.
In fact, careful analysis~\cite{Hou:2019uxa} suggest $B \to \tau\nu$ to be SM-like,
but $B \to \mu\nu$ probes the unique FCNH product 
$\rho_{tu}\rho_{\tau\mu}$ (Eq.~(\ref{eq:Bmunu-probe})), 
with $\rho_{tu}$ bringing in $V_{tb}/V_{ub}$ enhancement,
and $\rho_{\mu\tau}$ entering through the unobserved neutrino.
The decay can be enhanced or suppressed.
As seen from the current allowed (green) range in Fig.~\ref{fig:LFV_sum},
the variation is not huge around SM expectation.
But measured precisely, any deviation~\cite{Chang:2017wpl} of 
${\cal B}(B \to \mu\nu)/{\cal B}(B \to \tau\nu)$ from
0.0045 would not only be BSM, but beyond~\cite{Hou:2019uxa} 2HDM\;II.

The four modes of $B_{s,\, d} \to \mu^+\mu^-$ and $B^- \to \mu^-\nu$, $\tau^-\nu$
have good chance to become legacy modes for LHCb/CMS and Belle~II,
while $\mu$FV and $\tau$FV pursuits should also be watched.
Not to be forgotten are $K^0$, $B_d^0$ and $B_s^0$ (and also $D^0$) 
meson mixings, as well as $b\to s\gamma$.
For $B_s$ mixing, the CPV phase would become a probe
of CPV in {\it G}2HDM, while $b \to s\gamma$ has quite
a few parameters entering, making phenomenological analysis challenging.
We clearly see a new flavor era emerging.

{\it \textbf{\boldmath Anecdote: Muon $g-2$ as Harbinger}.---} \
The new FNAL muon $g-2$ result~\cite{Muong-2:2021ojo} shows that 
the experimental side is now confirmed, and would unfold further as data accumulates.
Although a new lattice study~\cite{Borsanyi:2020mff} casts 
some doubt on the ``theory consensus'' SM prediction~\cite{Aoyama:2020ynm},
hence on the 4.2$\sigma$ significance~\cite{Muong-2:2021ojo},
one should take the discrepancy seriously at present.

There is a known one-loop mechanism~\cite{Davidson:2010xv, Omura:2015nja} 
to account for muon $g-2$ in 2HDM\;III, i.e.\;{\it G}2HDM. 
With either $H$ or $A$ at 300~GeV in mass, 
we employ~\cite{Hou:2021sfl} large LFV Weinberg coupling 
$\rho_{\tau\mu} = \rho_{\mu\tau} \sim 0.2$ ($\sim 20\lambda_\tau$) 
to account for the muon $g-2$ anomaly.
But one needs to be close to the alignment limit of very small $h$--$H$ mixing 
to evade $h \to \tau\mu$ bound, Eq.~(\ref{eq:htamu}).
This motivates the check with $gg \to H,\,A \to \tau\mu$ search~\cite{Hou:2019grj}.
We find~\cite{Hou:2021sfl} that a recent result from CMS~\cite{CMS:2019pex}, 
using 36~fb$^{-1}$ data at 13~TeV, already puts more stringent bound 
on the extra top Weinberg coupling $\rho_{tt}$ than 
from $\tau \to \mu\gamma$ through the two-loop mechanism,
which is quite interesting.

The stringent constraint on $\rho_{tt}$ can be eased by
allowing the second extra top Weinberg coupling, $\rho_{tc}$,
to dilute the $H/A \to \tau\mu$ decay branching ratio, 
as the mass is below $t\bar t$ threshold.
As $H^+$ is taken as heavier, this then motivates~\cite{Hou:2021sfl}
the search for the novel signature of $cg \to bH^+ \to \tau^\pm\mu^\mp bW^+$,
as well as $cg \to bH^+ \to t\bar c bW^+$
 (same-sign dilepton plus two $b$-jets and additional jet, with missing $p_T$) at the LHC, 
on top of possible same-sign top plus jet signatures from $cg \to tH/A \to tt\bar c$.

The one-loop mechanism for solving muon $g-2$ with large $\rho_{\tau\mu}$
has strong repercussions on flavor physics, in particular muon-related physics, 
and may harbinger~\cite{Hou:2021qmf} a new $\mu/\tau$ era.
As discussed earlier, a similar one-loop diagram for $\mu \to e\gamma$
with $\tau$ in the loop, the enhanced $\rho_{\tau\mu}$ could
allow~\cite{Hou:2021qmf} even $\rho_{\tau e} \sim \lambda_e$ to bring the rate
right into the sensitivity of MEG~II! It also further elevates $\mu N \to eN$,
making it a potentially superb probe of $\rho_{qq}$.
By same token, $\tau \to \mu\gamma$ can probe $\rho_{\tau\tau}$
to below $\lambda_\tau$ strength, and
$\tau \to 3\mu$ can probe $\rho_{\mu\mu}$ to ${\cal O}(\lambda_\mu)$.

Thus, if large $\rho_{\tau\mu}$ is behind muon $g-2$ anomaly,
there can be strong implications for collider and flavor physics.
For example, we might soon see indications for $gg \to H, A \to \tau\mu$ 
at the LHC with full Run~2 data, 
or MEG\;II could discover $\mu \to e\gamma$ soon.
Although $\rho_{\tau\mu} \simeq 20\lambda_\tau$
grossly violates our conservative Eq.~(\ref{eq:rho3jrhoi1}),
it is up to Nature's choice, and maybe that is what 
she's been trying to tell us through the muon $g-2$ anomaly for two decades.
But we think that Nature cannot whimsically
move $\rho_{ij}^f$ elements around at random,
for our many sensitive flavor probes would have picked them up.

\section{Conclusion: Decadal Mission}

The discovery of the Higgs boson $h(125)$ is certainly
{\it the} landmark event at the Large Hadron Collider. 
But, after a dozen years running, a second ``discovery'' 
might be called {\it No New Physics} ({\it NNP}).

LHC certainly has made tremendous progress, in detailed measurements
and in pushing down on many search bounds.
But with {\it NNP} in sight,
HEP seems rudderless. Putting it differently,
there may now be too many little rudders and engines.
For instance, with the Weakly Interacting Massive Particle paradigm
strongly cornered, the bandwidth for Dark Matter (DM) search
has turned practically infinite, spanning 
from infinitesimal ``wave'' DM to massive black holes.
Thus, there are many out-of-the-box ideas and searches
for {\it Long-Lived Particles} (LLPs),
and one strong candidate, the Axion,
even spins off {\it Axion-Like Particle} (ALP) searches.
But given the ``infinite'' bandwidth, like shooting in the dark,
any given approach hitting target has very low likelihood.
The spreading-thin of manpower is not in the ``targeted'' tradition of HEP.

Another trend is to accept that there are no new particles 
below some scale $\Lambda$, then make an expansion in $1/\Lambda$.
This effective field theory (EFT) approach is not unreasonable,
and has gained in popularity by noting {\it NNP}.
But, have we exhausted all possibilities of dimension-4 
(Lagrandian) operators?

In this Invited Review, we have outlined what is in our mind
``{\it a most-likely New Physics in plain sight}'',
but most in HEP {\it See Not}: {a 2$^{nd}$ Higgs doublet} 
--- with two sets of dimension-4/Lagrangian couplings ---
i.e. {\it Extra Weinberg couplings} $\rho_{ij}^f$ ($f = u, d, \ell$),
and {\it Higgs quartic couplings} $\eta_i$ ($i = 1$--7). Impact: 
\begin{itemize}
\item Heavens:  $\eta_i$s at ${\cal O}(1)$ drive first order EWPT;

 \quad\quad\quad\quad\ $\rho_{tt}$ (or $\rho_{tc}$) drives Baryogenesis;

 Earth: \,\,\,\;\ $\rho_{ee}$ helps evade ACME'18 {$d_e$} bound.\footnote{
 Godspeed the success of ACME and similar experiments!
}
\item Sub-TeV $H, A, H^+$,  effects {\it hidden} so far by:

\quad\ \ \ {\it Flavor} structure (mass-mixing hierarchies);

\ \ \ \& emergent {\it Alignment} (small $h$-$H$ mixing).
\item Proposed:

\quad\ \ \,\ Direct Search Modes at LHC;

\ \ \ \& Myriad Indirect Flavor Probes.
\item Bonus: \,Landau pole at 10--20 TeV \,$\Rightarrow$\, New Scale.
\end{itemize}

\

In summary, we have before us the

\vskip0.2cm

\centerline{\underline{\underline{\bf Decadal Mission}}}
\vskip0.1cm
\centerline{
\underline{Find\;the\,$H, A, H^+$\,bosons\;and\;crack\;the\,{\it Flavor}\,\,code!}}


\vskip0.25cm

\centerline{\;\,The Mission would take several decades to unfold.}

\vskip0.4cm

\noindent{\bf Acknowledgments}\;
This work is the inaugural vision statement
of the new Academic Summit Project, MOST 110-2639-M-002-002-ASP
 (and the termination of MOST 109-2112-M-002-015-MY3) of Taiwan.
It is supported also by NTU grants 110L104019 and 110L892101.
We thank R. Jain and G. Kumar for reading the manuscript.



\begin{thebibliography}{0}%
\makeatletter
\providecommand \@ifxundefined [1]{%
 \@ifx{#1\undefined}
}%
\providecommand \@ifnum [1]{%
 \ifnum #1\expandafter \@firstoftwo
 \else \expandafter \@secondoftwo
 \fi
}%
\providecommand \@ifx [1]{%
 \ifx #1\expandafter \@firstoftwo
 \else \expandafter \@secondoftwo
 \fi
}%
\providecommand \natexlab [1]{#1}%
\providecommand \enquote  [1]{``#1''}%
\providecommand \bibnamefont  [1]{#1}%
\providecommand \bibfnamefont [1]{#1}%
\providecommand \citenamefont [1]{#1}%
\providecommand \href@noop [0]{\@secondoftwo}%
\providecommand \href [0]{\begingroup \@sanitize@url \@href}%
\providecommand \@href[1]{\@@startlink{#1}\@@href}%
\providecommand \@@href[1]{\endgroup#1\@@endlink}%
\providecommand \@sanitize@url [0]{\catcode `\\12\catcode `\$12\catcode
  `\&12\catcode `\#12\catcode `\^12\catcode `\_12\catcode `\%12\relax}%
\providecommand \@@startlink[1]{}%
\providecommand \@@endlink[0]{}%
\providecommand \url  [0]{\begingroup\@sanitize@url \@url }%
\providecommand \@url [1]{\endgroup\@href {#1}{\urlprefix }}%
\providecommand \urlprefix  [0]{URL }%
\providecommand \Eprint [0]{\href }%
\providecommand \doibase [0]{http://dx.doi.org/}%
\providecommand \selectlanguage [0]{\@gobble}%
\providecommand \bibinfo  [0]{\@secondoftwo}%
\providecommand \bibfield  [0]{\@secondoftwo}%
\providecommand \translation [1]{[#1]}%
\providecommand \BibitemOpen [0]{}%
\providecommand \bibitemStop [0]{}%
\providecommand \bibitemNoStop [0]{.\EOS\space}%
\providecommand \EOS [0]{\spacefactor3000\relax}%
\providecommand \BibitemShut  [1]{\csname bibitem#1\endcsname}%
\let\auto@bib@innerbib\@empty
\end{thebibliography}%


\begin{thebibliography}{99}

%
\bibitem{Anderson:1933mb}
  C.D.~Anderson,
  Phys. Rev. \textbf{43}, 491 (1933).
%
\bibitem{Neddermeyer:1937md}
  S.H.~Neddermeyer and C.D.~Anderson,
  Phys. Rev. \textbf{51}, 884 (1937).
%
\bibitem{Lattes:1947mx}
  C.M.G.~Lattes, G.P.S.~Occhialini and C.F.~Powell,
  Nature \textbf{160}, 453 (1947).
%
\bibitem{Hincks:1948vr}
  E.P.~Hincks and B.~Pontecorvo,
  Phys. Rev. \textbf{73}, 257 (1948).
%
\bibitem{Christenson:1964fg}
  J.H.~Christenson, J.W.~Cronin, V.L.~Fitch and R.~Turlay,
  Phys. Rev. Lett. \textbf{13}, 138 (1964).
%
\bibitem{Penzias:1965wn}
  A.A.~Penzias and R.W.~Wilson,
  Astrophys. J. \textbf{142}, 419 (1965).
%
\bibitem{Sakharov:1967dj}
  A.D.~Sakharov,
  Pisma Zh. Eksp. Teor. Fiz. \textbf{5}, 32 (1967).
%
\bibitem{Glashow:1961tr}
  S.L.~Glashow,
  Nucl. Phys. \textbf{22}, 579 (1961).
%
\bibitem{Meissner:1933}
  W.~Mei\ss ner and R. Ochsenfeld,
  Naturwiss. \textbf{21}, 787 (1933).
%
\bibitem{Anderson:1962}
  P.W.~Anderson,
  Phys. Rev. \textbf{130}, 439 (1962).
%
\bibitem{Englert:1964et}
  F.~Englert and R.~Brout,
  Phys. Rev. Lett. \textbf{13}, 321 (1964).
%
\bibitem{Higgs:1964pj}
  P.W.~Higgs,
  Phys. Rev. Lett. \textbf{13}, 508 (1964).
%
\bibitem{ATLAS:2012yve}
  G.~Aad \textit{et al.} [ATLAS],
  Phys. Lett. B \textbf{716}, 1 (2012).
%
\bibitem{CMS:2012qbp}
  S.~Chatrchyan \textit{et al.} [CMS],
  Phys. Lett. B \textbf{716}, 301 (2012).
%
\bibitem{Weinberg:1967tq}
  S.~Weinberg,
  Phys. Rev. Lett. \textbf{19}, 1264 (1967).
%
\bibitem{ATLAS:2016neq}
  G.~Aad \textit{et al.} [ATLAS and CMS],
  JHEP \textbf{08}, 045 (2016).
%
\bibitem{CMS:2020xwi}
  A.M.~Sirunyan \textit{et al.} [CMS],
  JHEP \textbf{01}, 148 (2021).
%
\bibitem{Hou:2012az}
  W.-S.~Hou,
  Chin. J. Phys. \textbf{50}, 375 (2012)
 [arXiv:1201.6029 [hep-ph]].
%
\bibitem{Kobayashi:1973fv}
  M.~Kobayashi and T.~Maskawa,
  Prog. Theor. Phys. \textbf{49}, 652 (1973).
%
\bibitem{Chang:2017wpl}
  P.~Chang, K.-F.~Chen and W.-S.~Hou,
  Prog. Part. Nucl. Phys. \textbf{97}, 261 (2017).
%
\bibitem{Wolfenstein:1983yz}
  L.~Wolfenstein,
  Phys. Rev. Lett. \textbf{51}, 1945 (1983).
%
\bibitem{BaBar:2001pki}
  B.~Aubert \textit{et al.} [BaBar],
  Phys. Rev. Lett. \textbf{87}, 091801 (2001).
%
\bibitem{Belle:2001zzw}
  K.~Abe \textit{et al.} [Belle],
  Phys. Rev. Lett. \textbf{87}, 091802 (2001).
%
\bibitem{Hou:2008xd}
  W.-S.~Hou,
  Chin. J. Phys. \textbf{47}, 134 (2009)
  [arXiv:0803.1234 [hep-ph]].
%
\bibitem{Grinstein:1987vj}
  B.~Grinstein, R.P.~Springer and M.B.~Wise,
  Phys. Lett. B \textbf{202}, 138 (1988).
%
\bibitem{Hou:1987kf}
  W.-S.~Hou and R.S.~Willey,
  Phys. Lett. B \textbf{202}, 591 (1988).
%
\bibitem{Glashow:1976nt}
  S.L.~Glashow and S.~Weinberg,
  Phys. Rev. D \textbf{15}, 1958 (1977).
%
\bibitem{CLEO:1994veu}
  M.S.~Alam \textit{et al.} [CLEO],
  Phys. Rev. Lett. \textbf{74}, 2885 (1995).
%
\bibitem{Hou:1992sy}
  W.S.~Hou,
  Phys. Rev. D \textbf{48}, 2342 (1993).
%
\bibitem{Grzadkowski:1991kb}
  B.~Grzadkowski and W.-S.~Hou,
  Phys. Lett. B \textbf{272}, 383 (1991).
%
\bibitem{ALEPH:1992zwu}
  D.~Buskulic \textit{et al.} [ALEPH],
  Phys. Lett. B \textbf{298}, 479 (1993).
%
\bibitem{Belle:2006but}
  K.~Ikado \textit{et al.} [Belle],
  Phys. Rev. Lett. \textbf{97}, 251802 (2006).
%
\bibitem{Fritzsch:1977za}
  H.~Fritzsch,
  Phys. Lett. B \textbf{70}, 436 (1977).
%
\bibitem{ARGUS:1987xtv}
  H.~Albrecht \textit{et al.} [ARGUS],
  Phys. Lett. B \textbf{192}, 245 (1987).
%
\bibitem{Cheng:1987rs}
  T.P.~Cheng and M.~Sher,
  Phys. Rev. D \textbf{35}, 3484 (1987).
%
\bibitem{Sher:1991yb}
  M.~Sher and Y.~Yuan,
  WM-91-112.
%
\bibitem{Hou:1988yu}
  W.-S.~Hou and R.G.~Stuart,
  Phys. Rev. Lett. \textbf{62}, 617 (1989).
%
\bibitem{Haeri:1988jt}
  B.~Haeri, A.~Soni and G.~Eilam,
  Phys. Rev. Lett. \textbf{62}, 719 (1989).
%
\bibitem{Hou:1990wz}
  W.-S.~Hou and R.G.~Stuart,
  Phys. Rev. D \textbf{43}, 3669 (1991).
%
\bibitem{Hou:1991un}
  W.-S.~Hou,
  Phys. Lett. B \textbf{296}, 179 (1992).
%
\bibitem{Chang:1993kw}
  D.~Chang, W.-S.~Hou and W.-Y.~Keung,
  Phys. Rev. D \textbf{48}, 217 (1993).
%
\bibitem{Harnik:2012pb}
  R.~Harnik, J.~Kopp and J.~Zupan,
  JHEP \textbf{03}, 026 (2013).
%
\bibitem{Hou:2020itz}
  W.-S.~Hou and G.~Kumar,
  Phys. Rev. D \textbf{102}, 115017 (2020).
%
\bibitem{Gell-Mann:1956iqa}
  M.~Gell-Mann,
  Nuovo Cim. \textbf{4}, 848 (1956) S2.
%
\bibitem{Kragh1907} 
  For a recent discussion, see H. Kragh,
  arXiv:1907.04623.
%
\bibitem{Glashow:1970gm}
  S.L.~Glashow, J.~Iliopoulos and L.~Maiani,
  Phys. Rev. D \textbf{2}, 1285 (1970).
%
\bibitem{Kane:2018oax}
  See e.g. G.~Kane,
  arXiv:1802.05199 [hep-ph].
%
\bibitem{PDG}
  P.A.~Zyla {\it et al.} [Particle Data Group],
  PTEP {\bf 2020}, 083C01 (2020).
%
\bibitem{ATLAS:2014lfm}
  G.~Aad \textit{et al.} [ATLAS],
  JHEP \textbf{06}, 008 (2014).
%
\bibitem{CMS:2014jkv}
  V.~Khachatryan \textit{et al.} [CMS],
  Phys. Rev. D \textbf{90}, 112013 (2014).
%
\bibitem{CMS:2015qee}
  V.~Khachatryan \textit{et al.} [CMS],
  Phys. Lett. B \textbf{749}, 337 (2015).
%
\bibitem{CMS:2021rsq}
  A.M.~Sirunyan \textit{et al.} [CMS],
  arXiv:2105.03007 [hep-ex].
%
\bibitem{Davidson:2005cw}
  S.~Davidson and H.E.~Haber,
  Phys. Rev. D \textbf{72}, 035004 (2005).
%
\bibitem{Hou:2019mve}
  W.-S.~Hou and T.~Modak,
  Phys. Rev. D \textbf{101}, 035007 (2020).
%
\bibitem{CMS:2021bdg}
  CMS Collaboration,
  CMS-PAS-TOP-20-007.
%
\bibitem{ACME:2018yjb}
  V.~Andreev \textit{et al.} [ACME],
  Nature \textbf{562}, 355 (2018).
%
\bibitem{Fuyuto:2017ewj}
  K.~Fuyuto, W.-S.~Hou and E.~Senaha,
  Phys. Lett. B \textbf{776}, 402 (2018)
%
\bibitem{Fuyuto:2019svr}
  K.~Fuyuto, W.-S.~Hou and E.~Senaha,
  Phys. Rev. D \textbf{101}, 011901(R) (2020).
%
\bibitem{Chiang:2016vgf}
  C.-W.~Chiang, K.~Fuyuto and E.~Senaha,
  Phys. Lett. B \textbf{762}, 315 (2016).
%
\bibitem{Altunkaynak:2015twa}
  B.~Altunkaynak, W.-S.~Hou, C.~Kao, M.~Kohda and B.~McCoy,
  Phys. Lett. B \textbf{751}, 135 (2015).
%
\bibitem{Planck:2013pxb}
  P.A.R.~Ade \textit{et al.} [Planck],
  Astron. Astrophys. \textbf{571}, A16 (2014).
%
\bibitem{ACME:2013pal}
  J.~Baron \textit{et al.} [ACME],
  Science \textbf{343}, 269 (2014).
%
\bibitem{Barr:1990vd}
  S.M.~Barr and A.~Zee,
  Phys. Rev. Lett. \textbf{65}, 21 (1990).
%
\bibitem{Cesarotti:2018huy}
  C.~Cesarotti, Q.~Lu, Y.~Nakai, A.~Parikh and M.~Reece,
  JHEP \textbf{05}, 059 (2019).
%
\bibitem{Cairncross:2017fip}
  W.B.~Cairncross {\it et al.},
  Phys. Rev. Lett. \textbf{119}, 153001 (2017).
%
\bibitem{Hou:2017hiw}
  W.-S.~Hou and M.~Kikuchi,
  EPL \textbf{123}, 11001 (2018).
%
\bibitem{Haber:2006ue}
  H.E.~Haber and D.~O'Neil,
  Phys. Rev. D \textbf{74}, 015018 (2006).
%
\bibitem{Haber:2010bw}
  H.E.~Haber and D.~O'Neil,
  Phys. Rev. D \textbf{83}, 055017 (2011).
%
\bibitem{Kanemura:2004ch}
  See e.g. S.~Kanemura, Y.~Okada and E.~Senaha,
  Phys. Lett. B \textbf{606}, 361 (2005).
%
\bibitem{Kohda:2017fkn}
  M.~Kohda, T.~Modak and W.-S.~Hou,
  Phys. Lett. B \textbf{776}, 379 (2018).
%
\bibitem{Barger:2010uw}
  V.~Barger, W.-Y.~Keung and B.~Yencho,
  Phys. Lett. B \textbf{687}, 70 (2010).
%
\bibitem{Hou:2019gpn}
  W.-S.~Hou, M.~Kohda and T.~Modak,
  Phys. Lett. B \textbf{798}, 134953 (2019).
%
\bibitem{CMS:2019rvj}
  A.M.~Sirunyan \textit{et al.} [CMS],
  Eur. Phys. J. C \textbf{80}, 75 (2020).
%
\bibitem{ATLAS:2021kqb}
  G.~Aad \textit{et al.} [ATLAS],
  arXiv:2106.11683 [hep-ex].
%
\bibitem{Belle:2019iji}
  M.T.~Prim, F.U.~Bernlochner, P.~ Goldenzweig, M.~Heck \textit{et al.} [Belle],
  Phys. Rev. D \textbf{101}, 032007 (2020).
%
\bibitem{Hou:2019uxa}
  W.-S.~Hou, M.~Kohda, T.~Modak and G.-G.~Wong,
  Phys. Lett. B \textbf{800}, 135105 (2020).
%
\bibitem{Ghosh:2019exx}
  D.K.~Ghosh, W.-S.~Hou and T.~Modak,
  Phys. Rev. Lett. \textbf{125}, 221801 (2020).
%
\bibitem{Hou:1997pm}
  W.-S.~Hou, G.-L.~Lin, C.-Y.~Ma and C.-P.~Yuan,
  Phys. Lett. B \textbf{409}, 344 (1997).
%
\bibitem{Hou:2018zmg}
  W.-S.~Hou, M.~Kohda and T.~Modak,
  Phys. Lett. B \textbf{786}, 212 (2018).
%
\bibitem{Hou:2019grj}
  W.-S.~Hou, R.~Jain, C.~Kao, M.~Kohda, B.~McCoy and A.~Soni,
  Phys. Lett. B \textbf{795}, 371 (2019).
%
\bibitem{Hou:2019qqi}
  W.-S.~Hou, M.~Kohda and T.~Modak,
  Phys. Rev. D \textbf{99}, 055046 (2019).
%
\bibitem{Hou:2020tnc}
  W.-S.~Hou, T.~Modak and T.~Plehn,
  SciPost Phys. \textbf{10}, 150 (2021).
%
\bibitem{Hou:2021xiq}
  W.-S.~Hou and T.~Modak,
  Phys. Rev. D \textbf{103}, 075015 (2021).
%
\bibitem{Hou:2020chc}
  W.-S.~Hou and T.~Modak,
  Mod. Phys. Lett. A \textbf{36}, 2130006 (2021).
%
\bibitem{Carena:2016npr}
  For a recent discussion, see M.~Carena and Z.~Liu,
  JHEP \textbf{11}, 159 (2016); and references therein.
%
\bibitem{CMS:2017ixp}
  A.M.~Sirunyan \textit{et al.} [CMS],
  Phys. Lett. B \textbf{778}, 349 (2018).
%
\bibitem{ATLAS:2017snw}
  M.~Aaboud \textit{et al.} [ATLAS],
  Phys. Rev. Lett. \textbf{119}, 191803 (2017).
%
\bibitem{CMS:2019lei}
  CMS Collaboration,
  CMS-PAS-HIG-17-027.
%
\bibitem{CMS:2019pzc}
  A.M.~Sirunyan \textit{et al.} [CMS],
  JHEP \textbf{04}, 171 (2020).
%
\bibitem{Hou:2018uvr}
  W.-S.~Hou, M.~Kohda and T.~Modak,
  Phys. Rev. D \textbf{98}, 075007 (2018).
%
\bibitem{CMS:2020imj}
  A.M.~Sirunyan \textit{et al.} [CMS],
  JHEP \textbf{07}, 126 (2020).
%
\bibitem{ATLAS:2021upq}
  G.~Aad \textit{et al.} [ATLAS],
  JHEP \textbf{06}, 145 (2021).
%
\bibitem{ATLAS:2020qdt}
 The ATLAS Collaboration,
  ATLAS-CONF-2020-027.
%
\bibitem{Bernon:2015qea}
  J.~Bernon, J.F.~Gunion, H.E.~Haber, Y.~Jiang and S.~Kraml,
  Phys. Rev. D \textbf{92}, 075004 (2015).
%
\bibitem{Bechtle:2016kui}
  P.~Bechtle, H.E.~Haber, S.~Heinemeyer, O.~St\r{a}l, T.~Stefaniak, G.~Weiglein and L.~Zeune,
  Eur. Phys. J. C \textbf{77}, 67 (2017).
%
\bibitem{Baak:2012kk}
  M.~Baak {\it et al.}, 
  Eur. Phys. J. C \textbf{72}, 2205 (2012).
%
\bibitem{BaBar:2012obs}
  J.P.~Lees \textit{et al.} [BaBar],
  Phys. Rev. Lett. \textbf{109}, 101802 (2012).
%
\bibitem{LHCb:2015gmp}
  R.~Aaij \textit{et al.} [LHCb],
  Phys. Rev. Lett. \textbf{115}, 111803 (2015).
%
\bibitem{LHCb:2015svh}
  R.~Aaij \textit{et al.} [LHCb],
  JHEP \textbf{02}, 104 (2016).
%
\bibitem{LHCb:2014vgu}
  R.~Aaij \textit{et al.} [LHCb],
  Phys. Rev. Lett. \textbf{113}, 151601 (2014).
%
\bibitem{LHCb:2017avl}
  R.~Aaij \textit{et al.} [LHCb],
  JHEP \textbf{08}, 055 (2017).
%
\bibitem{Hou:2019dgh}
  G.W.-S.~Hou,
  Int. J. Mod. Phys. A \textbf{34}, 1930002 (2019).
%
\bibitem{Muong-2:2021ojo}
  B.~Abi \textit{et al.} [Muon g-2],
  Phys. Rev. Lett. \textbf{126}, 141801 (2021).
%
\bibitem{Muong-2:2006rrc}
  G.W.~Bennett \textit{et al.} [Muon g-2],
  Phys. Rev. D \textbf{73}, 072003 (2006).
%
\bibitem{Bordone:2017bld}
  M.~Bordone, C.~Cornella, J.~Fuentes-Mart\'\i{}n and G.~Isidori,
  Phys. Lett. B \textbf{779}, 317 (2018).
%
\bibitem{Pati:1974yy}
  J.C.~Pati and A.~Salam,
  Phys. Rev. D \textbf{10}, 275 (1974).
%
\bibitem{Belle-II:2018jsg}
  E.~Kou, P.~Urquijo \textit{et al.} [Belle-II],
  PTEP \textbf{2019}, 123C01 (2019).
%
\bibitem{Belle:2021ysv}
  K.~Uno, K.~Hayasaka, K.~Inami \textit{et al.} [Belle],
  arXiv:2103.12994 [hep-ex].
%
\bibitem{Bordone:2018nbg}
  M.~Bordone, C.~Cornella, J.~Fuentes-Mart\'\i{}n and G.~Isidori,
  JHEP \textbf{10}, 148 (2018).
%
\bibitem{Glashow:2014iga}
  S.L.~Glashow, D.~Guadagnoli and K.~Lane,
  Phys. Rev. Lett. \textbf{114}, 091801 (2015).
%
\bibitem{LHCb:2019ujz}
  R.~Aaij \textit{et al.} [LHCb],
  Phys. Rev. Lett. \textbf{123}, 211801 (2019).
%
\bibitem{Cornella:2019hct}
  C.~Cornella, J.~Fuentes-Mar\'\i{}n and G.~Isidori,
  JHEP \textbf{07}, 168 (2019).
%
\bibitem{Belle:2021rod}
  H.~Atmacan, A.J. ~Schwartz, K.~Kinoshita \textit{et al.} [Belle],
  arXiv:2108.11649 [hep-ex].
%
\bibitem{BaBar:2012azg}
  J.P.~Lees \textit{et al.} [BaBar],
  Phys. Rev. D \textbf{86}, 012004 (2012).
%
\bibitem{LHCb:2020khb}
  R.~Aaij \textit{et al.} [LHCb],
  JHEP \textbf{06}, 129 (2020).
%
\bibitem{Borsanyi:2020mff}
  S.~Borsanyi {\it et al.}, 
  Nature \textbf{593}, 51 (2021)
  [arXiv:2002.12347 [hep-lat]].
%
\bibitem{Aoyama:2020ynm}
  T.~Aoyama {\it et al.}, 
  Phys. Rept. \textbf{887}, 1 (2020).
%
\bibitem{Davidson:2010xv}
  S.~Davidson and G.J.~Grenier,
  Phys. Rev. D \textbf{81}, 095016 (2010).
%
\bibitem{Omura:2015nja}
  Y.~Omura, E.~Senaha and K.~Tobe,
  JHEP \textbf{05}, 028 (2015).
%
\bibitem{Hou:2021sfl}
  W.-S.~Hou, R.~Jain, C.~Kao, G.~Kumar and T.~Modak,
  arXiv:2105.11315 [hep-ph].
%
\bibitem{CMS:2019pex}
  A.M.~Sirunyan \textit{et al.} [CMS],
  JHEP \textbf{03}, 103 (2020).
%
\bibitem{Hou:2021qmf}
  W.-S.~Hou and G.~Kumar,
  arXiv:2107.14114 [hep-ph].

\end{thebibliography}
\end{document}